\documentstyle[epsfig,macros]{lamuphys}
\makeatletter
\let\chapter\hid@chapter
\makeatother
\begin{document}
\pagenumbering{arabic}
\title{
Non-perturbative renormalization of QCD$^1$
}

\author{Rainer Sommer 
}

\institute{DESY-IfH, Platanenallee 6, D-15738 Zeuthen}

\maketitle
\vspace{-4.5cm} 


{ \normalsize
\hfill \parbox{21mm}{DESY 97-207}}\\[40mm]
\addtocounter{footnote}{1}
\footnotetext{
Lectures given at the {\it 36. Internationale
Universit\"atswochen f\"ur Kern- und Teilchenphysik (Schladming  1997) : Computing Particles}}

\begin{abstract}
In these lectures, we discuss different types of renormalization problems
in QCD and their non-perturbative solution in the framework of
the lattice formulation. In particular the 
recursive finite size methods to compute the scale-dependence
of renormalized quantities is explained. An important ingredient
in the practical applications is  
the Schr\"odinger functional. It is introduced and its renormalization
properties are discussed. \\
Concerning applications, the computation of the running coupling 
and the running quark mass
are covered in detail and it is shown how the $\Lambda$-parameter and
renormalization group invariant quark mass can be obtained.
Further topics are the renormalization of isovector currents and 
non-perturbative Symanzik improvement.\\[1ex]
{\bf Contents}
\begin{itemize}
\item[1.]{ Introduction\\ \it
Basic renormalization: hadron spectrum; 
Finite renormalization: (semi-)leptonic decays;
Scale dependent renormalization;
Irrelevant operators}
\item[2.]{The problem of scale dependent renormalization\\ \it
The extraction of $\alpha$ from experiments;
Reaching large scales in lattice QCD}
\item[3.]{ The Schr\"odinger functional\\ \it
Definition;
Quantum mechanical interpretation;
Background field;
Perturbative expansion;
General renormalization properties;
Renormalized coupling;
Quarks;
Renormalized mass;
Lattice formulation}
\item[4.]{The computation of $\alpha(q)$\\ \it
The step scaling function;
Lattice spacing effects in perturbation theory;
The continuum limit -- universality;
The running of the coupling;
The low energy scale; 
Matching at finite energy;
The $\Lambda$ parameter of quenched QCD;
The use of bare couplings}
\item[5.]{Renormalization group invariant quark mass}
\item[6.]{Chiral symmetry, normalization of currents and $\Oa$-improvement\\ \it
Chiral Ward identities;
$\Oa$-improvement;
Normalization of isovector currents}
\item[7.]{Summary, Conclusions}
\end{itemize}

\end{abstract}
%

\section{Introduction}
The topic of
these lectures is the   
computation of properties of particles that are bound by the strong
interaction  or more generally interact strongly. The strong interactions 
are theoretically described by Quantum Chromo Dynamics (QCD), a local 
quantum field theory. 

Starting from the Lagrangian of a field theory, predictions for cross sections
and other observables are usually made by applying renormalized perturbation 
theory, the expansion in terms of the (running) couplings of the theory.
While this expansion is well controlled as far as electroweak interactions
are concerned, its application in QCD is limited to high energy processes where
the QCD coupling, $\alpha$, is sufficiently
small. In general -- and in particular for the calculation of bound state 
properties -- a non-perturbative solution of the theory is required.

The only method that is known to address this problem is the 
numerical simulation of the 
Euclidean path integral of QCD on a space-time lattice.  
By ``solution of the theory'' we here mean that one poses a
well defined question like ``what is the value of the $\pi$ decay constant'',
and obtains the answer (within a certain precision) through a series of
Monte Carlo (MC) simulations. This then allows to test the agreement of theory
and experiment on the one hand and helps in the determination
of Standard Model parameters from experiments on the other hand.

Quantum field theories are defined by first formulating them 
in a regularization with an
ultraviolet cutoff $\Lambda_{\rm cut}$ and then considering the limit  
$\Lambda_{\rm cut} \to \infty$. In the lattice formulation~(\cite{Wilson}), 
the cutoff
is given by the inverse of the lattice spacing $a$;
we have to consider the continuum limit $a \to 0$. 
At a finite value of $a$, the theory is defined in terms of the bare coupling constant, bare masses
and bare fields. Before making predictions for experimental observables 
(or more generally for observables that have a well defined continuum
limit) the coupling, masses and fields have to be renormalized.
This is the subject of my lectures. 

Renormalization is an ultraviolet phenomenon with relevant momentum
scales of order $a^{-1}$. Since $\alpha$
becomes weak in the ultraviolet, one expects to be able to perform 
renormalizations 
perturbatively, i.e.
computed in a power series in $\alpha$ as one approaches the continuum 
limit $a \to 0$.\footnote{
%
%
For simplicity we ignore here the cases of mixing of a given operator 
with operators of lower dimension where this statement does not hold.}  
However, one has to take care about the following point.
In order to keep the numerical effort of a simulation tractable,
the number of degrees of freedom in the simulation may not be excessively 
large. This means that  the lattice spacing $a$ can not be taken very much smaller than the
relevant physical length scales of the observable that is considered.
Consequently the momentum scale $a^{-1}$ that is relevant for the renormalization is not always large enough to justify the truncation of the perturbative series. 
In order to obtain a truly non-perturbative answer, the renormalizations have to be performed non-perturbatively. 

Depending on the observable,
the necessary renormalizations are of different nature. I will use
this introduction to point out the different types 
and in particular explain the problem that occurs in a non-perturbative
treatment of renormalization.

\subsection{Basic renormalization: hadron spectrum\label{s_hs}}
At this school, the calculation of the hadron spectrum is 
covered in detail in the lectures of Don Weingarten~\myref{Don}. I mention it
anyway because I want to make the conceptual point that it can be
considered as a non-perturbative renormalization. I refer the reader 
to Weingarten's lectures both for details in such calculations
and for an introduction to the basics of
lattice QCD. 

The calculation starts by choosing certain values for the bare coupling,
$g_0$, and the bare masses of the quarks in units of the lattice spacing,
$a \mbare^f$. The flavor index $f$ assumes values $f={\rm u,d,s,c,b}$ for the 
up, down, charm and bottom quarks that are sufficient to describe
hadrons of up to a few $\GeV$ masses.
We neglect isospin breaking and take 
the light quarks to be degenerate, 
$\mbare^{\rm u}=\mbare^{\rm d}=\mbare^{\rm l}$.

Next, from MC simulations of suitable correlation functions, one computes 
masses of five different hadrons $H$, e.g. $H={\rm p},\pi,{\rm K,D,B}$
for the proton, the pion and the K-,D- and B-mesons,
\bes
  a m_H = a m_H(g_0,a \mbare^{\rm l}, a \mbare^{\rm s}, 
                 a \mbare^{\rm c},a \mbare^{\rm b}) \enspace .      \label{hadrons}
\ees
The theory is renormalized by first
setting $m_{\rm p}=m_{\rm p}^{\rm exp}$, where 
$m_{\rm p}^{\rm exp}$ is the experimental value of the proton mass. This determines the lattice spacing via
\bes
  a= (a m_{\rm p}) / {m_{\rm p}^{\rm exp}} \enspace .      \label{spacing}
\ees
Next one must 
choose the parameters $a \mbare^{f}$ such that (\ref{hadrons}) is 
indeed satisfied
with the experimental values of the meson masses. Equivalently, one
may say that at a given value of $g_0$ one fixes the bare quark masses
from the condition
\bes 
(a m_H) / (a m_{\rm p}) = m_H^{\rm exp} / m_{\rm p}^{\rm exp} \, , \quad 
       H=\pi,{\rm K,D,B} 
                                           \enspace .      \label{mesons}
\ees
and the bare coupling $g_0$ then determines the value
of the lattice spacing through \eq{spacing}.

After this {\it renormalization}, namely {\it the elimination of the bare parameters
in favor of physical observables}, the theory is completely defined and
predictions may be made. E.g. the mass of the $\Delta$-resonance can be
determined,
\bes 
 m_{\Delta}=a^{-1} [am_\Delta][1 + \Oa] \enspace .      \label{e_Delta}
\ees 
For the rest of this section, I assume that the bare parameters
have been eliminated  and consider the additional renormalizations
of more complicated observables.

\paragraph{Note.} Renormalization as described here is done without any reference to perturbation theory.
One could in principle use the perturbative formula for $(a \Lambda)(g_0)$
for the renormalization of the bare coupling, where $\Lambda$ denotes the
$\Lambda$-parameter of the theory. Proceeding in this way, one obtains
a further prediction namely  $m_{\rm p}/\Lambda$ but at the price of 
introducing $\rmO(g_0^2)$ errors in the prediction of the observables.
As mentioned before, such errors decrease very slowly as one performs 
the continuum limit. A better method to compute the $\Lambda$-parameter
will be discussed later. 
\subsection{Finite renormalization: (semi-)leptonic decays\label{s_finiter}}
Semileptonic weak decays
of hadrons such as ${\rm K} \to \pi \,\, e \,\, \bar{\nu}$ 
are mediated by electroweak 
vector bosons. These couple to quarks through linear 
combinations of vector and axial vector flavor currents. 
Treating the electroweak interactions at 
lowest order, the decay rates are given in terms of QCD  
matrix elements of these currents. For simplicity we consider only
two flavors; an application is then the computation of the pion decay 
constant describing the leptonic decay 
$\pi \to  e \,\, \bar{\nu}$.\footnote{
Of course, decays of hadrons containing b-quarks are more interesting phenomenologically, but here our emphasis is on the principle of 
renormalization.}
The currents are
\bes            
  A_{\mu}^a(x) &=& 
            \psibar(x)\dirac{\mu}\dirac{5}\frac{1}{2}\tau^a\psi(x) \enspace ,
            \nonumber \\
  V_{\mu}^a(x) &=& 
            \psibar(x)\dirac{\mu}\frac{1}{2}\tau^a\psi(x) \enspace ,
            \label{e_currents} 
\ees
where $\tau^a$ denote the Pauli matrices which act on the flavor indices of
the quark fields.
A priori the bare currents \eq{e_currents} need renormalization. 
However, in the limit of vanishing quark masses the (formal continuum) QCD Lagrangian is invariant
under SU$(2)_{\rm V} \times $ SU$(2)_{\rm A}$ flavor symmetry transformations.
This leads to nonlinear relations between the currents called
current algebra, from which one concludes that no renormalization is 
necessary (cf. \sect{s_curr}).

In the regularized theory SU$(2)_{\rm V} \times $ SU$(2)_{\rm A}$ is not
an exact symmetry but is violated by terms of order $a$.
As a consequence there is a  finite
renormalization~\myref{Za1,Za2,Za3,Za4,Za5}
\bes
 (\ar)_{\mu}^a &=& \za A_{\mu}^a \enspace ,\nonumber \\
 (\vr)_{\mu}^a &=& \zv V_{\mu}^a \enspace , \label{e_zazv}
\ees
with renormalization constants
$\za,\zv$ that do not contain any logarithmic (in $a$)
or power law divergences and do not depend on any
physical scale. Rather they are approximated by
\bes
  \za = 1 + \za^{(1)} g_0^2 + \ldots \enspace ,\nonumber \\
  \zv = 1 + \zv^{(1)} g_0^2 + \ldots \enspace ,\label{e_zazv_pert} 
\ees
for small $g_0$.

On the non-perturbative level these renormalizations  
can be fixed by current algebra relations \myref{Boch,MaMa,paper4}
as will be explained in section~\ref{s_curr}. 

\subsection{Scale dependent renormalization \label{s_sdr}}
\subsubsection{a) Short distance parameters of QCD.}
As we take the relevant length scales in correlation functions 
to be small  or take the energy scale in 
scattering processes to be high, QCD becomes a theory of weakly 
coupled quarks and gluons. The strength of the interaction may be 
measured for instance by the ratio of 
the production rate of three jets to the rate for two jets in high energy
$e^+ ~ e^-$ collisions,
\bes
\alpha(q) &\propto& 
 {\sigma(e^+ ~ e^- \to q ~ \bar{q} ~g) \over \sigma(e^+ ~ e^- \to q ~ \bar{q})}
  \, , \quad q^2=(p_{e^-}+p_{e^+})^2 \gg  10 \GeV^2 \enspace . \label{e_jets}
\ees
We observe the following points. 
  \begin{itemize}
     \item{The perturbative renormalization group tells us that $\alpha(q)$       
          decreases logarithmically with growing energy $q$. In other words
           the renormalization from the bare coupling to a renormalized one
           is logarithmically scale dependent.} 
     \item{Different definitions of $\alpha$ are possible; but with 
           increasing energy, $\alpha$ 
           depends less and less on the definition (or the process).} 
     \item{In the same way, running quark masses $\mbar$ acquire a precise       
             meaning
           at high energies.}    
     \item{Using a suitable definition (scheme), the $q$-dependence of 
           $\alpha$ and
           $\mbar$ can be determined non-perturbatively and at high energies
           the short distance parameters $\alpha$ and $\mbar$ can be converted
            to any other scheme using perturbation theory in $\alpha$.}
  \end{itemize}
Explaining these points in detail is the main objective of my lectures.
For now we proceed to give a second example of scale dependent renormalization.

\subsubsection{b) Weak hadronic matrix elements of 4-quark operators.}
Another example of scale dependent renormalization
is the 4-fermion operator, $O^{\Delta_s=2}$, which changes strangeness by 
two units. It originates
from weak interactions after integrating out the fields that have high masses.
It describes the famous mixing in the neutral Kaon system 
through the
matrix element
$$\langle\overline{\rm K}^{0}| O^{\Delta_s=2}(\mu) | 
           {\rm K}^{0}\rangle \enspace .$$
Here the operator renormalized at energy scale $\mu$
is  given by 
\bes
O^{\Delta_s=2}(\mu) &=& Z^{\Delta_s=2} (\mu a,g_0)
 \left\{ {\psibars} \gamma_\mu^{\rm L} \psid \, {\psibars} \gamma_\mu^{\rm L}              \psid +
 \sum_{j={\rm S,P,V,A,T}}                                     
 z_j {\psibars} \Gamma_j \psid \, {\psibars} \Gamma_j \psid \right\}  
       \enspace , \nonumber \\
  && \gamma_\mu^{\rm L}=\frac12 \gamma_\mu(1-\gamma_5)
    \enspace , \nonumber \\
 && \Gamma_{\rm S} = 1, \,\Gamma_{\rm P} = \gamma_5,\, \ldots,
    \Gamma_{\rm T} = \sigma_{\mu \nu}                      
    \enspace ,    \nonumber \\
 && z_j  = {\rm O}(g_0^2),\quad z_{\rm V}=-z_{\rm A}  \label{e_opkk}
\ees
where I have indicated the flavor index of the quarks explicitly.
A mixing of the leading bare operator, 
${\psibars} \gamma_\mu^{\rm L} \psid \, {\psibars} \gamma_\mu^{\rm L} \psid$,
with operators of different chirality is again possible since the 
lattice theory does not have an exact chiral symmetry
for finite values of the lattice spacing. The mixing coefficients
$z_j$ may be fixed non-perturbatively by current algebra~\myref{Wardid_jap}. 
Afterwards, the overall
scale dependent renormalization has to be treated in the same way as
the renormalization of the coupling. 

\subsection{Irrelevant operators \label{s_Io}}
A last category of renormalization is associated with the removal of 
lattice discretization errors such as the $\Oa$-term in \eq{e_Delta}.
Following Symanzik's improvement program, this can be achieved order by order
in the lattice spacing by adding irrelevant operators, i.e. operators
of dimension larger than four, to the lattice Lagrangian~\myref{Symanzik}. 
The coefficients
of these operators are easily determined at tree level of perturbation theory,
but in general they need to be renormalized.

In this subject significant progress
has been made recently as reviewed by \cite{Lepage96,Sommer97}. 
In particular the latter reference 
is concerned with non-perturbative Symanzik 
improvement and uses a notation consistent with the one 
of these lectures. It will become evident in later sections
that improvement is very
important for the progress in lattice QCD.

Note also the alternative approach of removing lattice artifacts 
order by order in the coupling constant but non-perturbatively in the lattice spacing $a$ as recently reviewed 
by \cite{Nieder}.

\section{The problem of scale dependent renormalization\label{s_p}}
Let us investigate the extraction of short distance parameters (Section \ref{s_sdr}a) in more detail.
First we analyze the conventional way of obtaining $\alpha$ from experiments.
Then we explain how one can compute $\alpha$ at large energy scales using lattice QCD.

\subsection{The extraction of $\alpha$ from experiments}
One considers experimental observables $O_i$ depending on an 
overall energy scale $q$ and possibly some additional kinematical 
variables denoted by $y$. 
The observables can be computed in a perturbative series 
which is usually written in terms of the $\MSbar$ coupling $\alphaMSbar$,
\footnote{We can always arrange the definition of the observables such that 
they start with a term $\alpha$.}
\bes
 O_i(q,y) &=& \alphaMSbar(q) + A_i(y)\alphaMSbar^2(q)+\ldots 
 \enspace . \label{e_O_i}
\ees
For example $O_i$ may be constructed from jet cross sections and
$y$ may be related to the details of the definition of a jet.

\begin{figure}
\vspace{-1.2cm}

\hspace{2cm}
\psfig{file=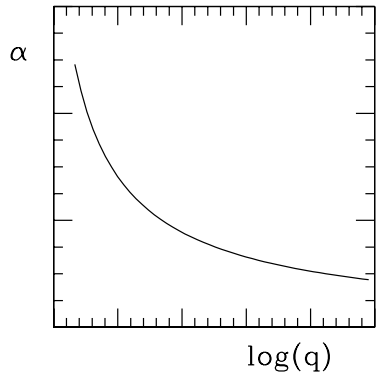, width=7cm}
\vspace{-0.5cm}
\caption{Running of $\alpha$ in a definite scheme.\label{f_running1}}
\end{figure}

The renormalization group describes the energy dependence
of $\alpha$ in a general scheme ($\alpha \equiv \gbar^2/(4\pi)$),
\bes
  q {\partial \bar g \over \partial q} &=& \beta(\bar g) \enspace ,
     \label{e_RG}
\ees
where the $\beta$-function has an asymptotic expansion     
\bes     
 \beta(\bar g) & \buildrel {\bar g}\rightarrow0\over\sim &
 -{\bar g}^3 \left\{ b_0 + {\bar g}^{2}  b_1 + \ldots \right\}
                      \enspace , \nonumber\\
 &&b_0=\frac{1}{(4\pi)^2}\bigl(11-\frac{2}{3}\nf\bigr) 
                      \enspace ,\nonumber\\
 &&b_1=\frac{1}{(4\pi)^4}\bigl(102-\frac{38}{3}\nf\bigr) \enspace , 
 \label{e_RGpert}
\ees
with higher order coefficients $b_i, \, i>1 $ that depend on the scheme.
\Eq{e_RGpert} entails the aforementioned property of asymptotic freedom:
at energies that are high enough for \eq{e_RGpert} to be applicable and
for a number of quark flavors, $\nf$, that is not too large, 
$\alpha$ decreases with increasing energy as indicated in \fig{f_running1}.
The asymptotic solution of \eq{e_RG} is given by
\bes
\gbar^2 & \buildrel {q}\rightarrow \infty \over\sim &
   {1 \over b_0 \ln(q^2/\Lambda^2)}
  -{b_1 \ln[\ln(q^2/\Lambda^2)] \over b_0^3 [\ln(q^2/\Lambda^2)]^2}
  + \rmO\left( {\{\ln[\ln(q^2/\Lambda^2)]\}^2 \over [\ln(q^2/\Lambda^2)]^{3}} \right)
\ees
with $\Lambda$ an integration constant which is different
in each scheme.

We note that -- neglecting experimental uncertainties -- $\alphaMSbar$ 
extracted in this way is obtained with a precision given by the terms
that are left out in \eq{e_O_i}. In addition to $\alpha^3$-terms, there
are non-perturbative contributions which may originate
from ``renormalons'', ``condensates'' (the two possibly being related),
``instantons'' or -- most importantly -- may have an origin that no physicist
has yet uncovered. Empirically, one observes that values
of $\alphaMSbar$ determined
at different energies and evolved to a common reference point 
using the renormalization group equation \eq{e_RG} including $b_2$
agree rather well with each other; the aforementioned uncertainties are 
apparently not very large.
Nevertheless, determinations of $\alpha$ are limited in precision
because of these uncertainties and in particular if there was a significant
discrepancy between $\alpha$ determined at different energies one would not
be able to say whether this was due to the terms left out in \eq{e_O_i}
or
was due to terms missing in the Standard Model Lagrangian, eg. 
an additional strongly interacting matter field.

It is an obvious possibility and at the same time a challenge for 
lattice QCD to achieve a determination of $\alpha$ in one (non-perturbatively) 
well defined scheme and evolve this coupling to high energies. There it may be 
used to compute jet cross sections and compare to high energy experiments
to test the agreement between theory and experiment. Since in the lattice
regularization QCD is naturally renormalized through the hadron spectrum,
such a calculation provides the connection between low energies
and high energies, verifying that one and the same theory describes 
both the hadron spectrum and the properties of jets. 

\paragraph{Note.} A dis-satisfying property of $\alphaMSbar$ is that it is 
{\it only} defined in a perturbative framework; strictly speaking there is
no meaning of phrases like ``non-perturbative corrections'' in the extraction 
of $\alphaMSbar$ from experiments. The way that I have written \eq{e_O_i} 
suggests immediately what should be done instead. An observable $O_i$ itself 
may be taken as a definition of $\alpha$ -- of course with due care. 
Such  schemes called {\it physical schemes} are defined
without ambiguities. 
This is what will be done below
for observables that are easily handled in MC-simulations of QCD.
For an additional example see ~\cite{Grunberg}.

\subsection{Reaching large scales in lattice QCD \label{s_Rls}}
Let us simplify the discussion and restrict ourselves to
 the pure Yang-Mills theory
without matter fields in this section.
A natural candidate for a non-perturbative definition of $\alpha$
is the following. Consider a quark and an anti-quark separated by a distance
$r$ and in the limit of infinite mass. They feel a force $F(r)$, 
the derivative of the static potential $V(r)$, which can be 
computed 
from Wilson loops~(see e.g. \cite{MM}). A physical coupling is defined as
\bes
 \alpha_{\rm q \bar q} (q) &\equiv& { \frac{1}{C_F} r^2 F(r) \, , \quad q=1/r,} 
 { \quad C_F = 4/3} \enspace . \label{e_alphaqq}
\ees
It is related to the $\MSbar$ coupling by
\bes
 \alpha_{\rm q \bar q} &=& \alphaMSbar \, + \, c_1^{\MSbar \,\rm q \bar q} \alphaMSbar^2 
 \, + \, \ldots \enspace , \label{e_alphaqq_pert}
\ees
where both couplings are taken at the same energy scale and
the coefficients in their perturbative relation are pure numbers.
The 1-loop coefficient, $c_1^{\MSbar\,\rm q \bar q}$, also determines
the ratio of the $\Lambda$-parameters vs.
\bes
\Lambda_{\rm q \bar q} / \Lambda_{\MSbar} =\exp(-c_1^{\MSbar\,\rm q \bar q}
                                           /(8\pi b_0))
\enspace . \label{e_lambda_ratio}
\ees
Note that $\alpha_{\rm q \bar q}$ is a renormalized coupling defined 
in continuum QCD.

\paragraph{Problem.}
If we want to achieve what was proposed in the previous subsection,
the following criteria must be met.
\begin{itemize}
 \item{Compute $\alpha_{\rm q \bar q} (q)$ at energy scales of 
       $q\sim 10\,\GeV$ or higher in order to be able to make the connection 
       to
       other schemes with controlled perturbative errors.}
 \item{Keep the energy scale $q$ removed from the cutoff $a^{-1}$ to
       avoid large discretization effects and to be able to 
       extrapolate to the continuum limit.} 
 \item{Of course, only a finite system can be simulated by MC.
       To avoid finite
       size effects one must keep the box size $L$ large
       compared to the confinement scale ${K}^{-1/2}$ to avoid finite
       size effects. Here, $K$ denotes the 
       string tension, $K=\lim_{r\to\infty}F(r)$.}       
\end{itemize}
These conditions are summarized by
\bes
  L \quad \gg \,\, { 1 \over 0.4\GeV} \,\, \gg \,\,
  {1 \over q} \, \sim \, {1 \over 10\GeV}  \,\,  \gg a \enspace ,
   \label{e_conditions}
\ees       
which means that one must perform a MC-computation of an $N^4$ lattice
with $N \equiv L/a \gg 25$. 
It is at present impossible to perform such a computation. 
The origin of this problem 
is simply that the extraction of short distance parameters
requires that one covers physical scales that are quite disparate.
To cover these scales in one simulation requires a very fine resolution,
which is too demanding for a MC-calculation.

Of course, one may attempt to compromise in various ways. E.g. one may 
perform phenomenological corrections for lattice artifacts,
keep $1/q \sim a$ and at the same time reduce the value of $q$ compared to 
what I quoted in \eq{e_conditions}.  Calculations of
$\alpha_{\rm q \bar q}$ along these lines have been 
performed in the Yang-Mills theory  \myref{Michael,UKQCDI,BaSch}. 
It is difficult
to estimate the uncertainties due to the approximations that are necessary 
in this approach. 

\paragraph{Solution.}
Fortunately these compromises can be avoided altogether 
\myref{alphaI}. 
The solution to the problem is to identify the two physical scales, above,
\bes
 q =1/L \enspace  .
\ees 
In other words, one
takes a finite size effect as the physical observable. The evolution of the coupling
with $q$ can then be computed in several steps, changing $q$ by factors of
order $2$ in each step. In this way, no large scale ratios appear
and discretization errors are  small for $L/a \gg 1$. 

For illustration, we modify the definition of $\alpha_{\rm q \bar q} (q)$
to fit into this class of finite volume couplings.
Consider the Yang-Mills theory on a $ T \times L^3$ -- torus with 
$T \gg L$.\footnote{
It is well known that perturbation theory in small volumes with periodic boundary conditions is complicated 
by the occurrence of zero modes 
\myref{metamo,Lu83}. These can be avoided by choosing
twisted periodic boundary conditions in 
space \myref{thooft,Baal,LuWe}.  
}
The finite volume coupling,
\bes
 \tilde{\alpha}_{\rm q \bar q}(q) &\equiv& k \{ r^2 F(r,L)\}_{r=L/4} \, , \quad q=1/L \enspace ,
\ees
can again be related to the $\MSbar$ coupling perturbatively,
\bes  
 \tilde{\alpha}_{\rm q \bar q} &=& \alphaMSbar + \tilde{c}_1^{\MSbar\,\rm q \bar q} 
                              \alphaMSbar^2 + \ldots \enspace .
\ees
This relation may come as a surprise since it relates 
a small volume quantity to an infinite volume one. Remember, however,
that once the bare coupling and masses are eliminated there are no free 
parameters. Renormalized couplings in finite volume and couplings in infinite
volume are in one-to-one correspondence. When they are small they can be 
related by perturbation theory. In particular, \eq{e_lambda_ratio}
holds with the obvious modification.

\begin{figure}[ht]
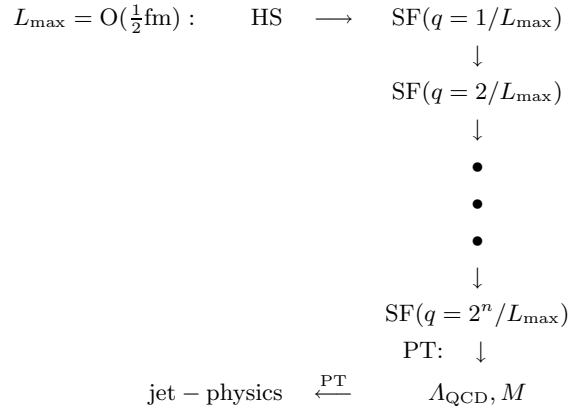

\bes
 { L_{\rm max}}=\rmO(\frac{1}{2}\fm): \qquad 
 {\rm HS} \quad \longrightarrow \quad
      &{\rm SF} (q=1/{ L_{\rm max}})& \quad 
               \nonumber \\
      &\downarrow&  \nonumber \\
      &{\rm SF} (q=2/{ L_{\rm max}})&  \nonumber \\ 
      &\downarrow&  \nonumber \\
      &\bullet&  \nonumber \\
      &\bullet&  \nonumber \\
      &\bullet&  \nonumber \\
      &\downarrow&  \nonumber \\
      &{\rm SF} (q=2^n/{ L_{\rm max}})& \nonumber \\
   &\mbox{ \small PT:} \quad  \downarrow \qquad \quad &  \nonumber \\
{\rm jet-physics} \quad \stackrel{\rm  PT}{\longleftarrow} 
     \quad   &\Lambda_{\rm QCD}, M & \nonumber
\ees
\vspace{-0.8cm}
\caption{The strategy for a non-perturbative computation of 
         short distance parameters.
\label{f_strategy}}
\end{figure}

The complete strategy to compute short distance parameters 
is summarized in \fig{f_strategy}.
One first renormalizes QCD replacing the bare parameters by hadronic
observables. This defines the hadronic scheme (HS) as explained in Sect.~\ref{s_hs}. At a low energy scale $q=1/L_{\rm max}$ this scheme 
can be
related to the finite volume scheme denoted by SF in the graph. 
Within this scheme one then computes the scale evolution
up to a desired energy $q=2^n/{ L_{\rm max}}$. As we will see it 
is no problem to choose the number of steps $n$ large enough to be
sure that one is in the perturbative regime. There 
perturbation theory (PT) is used to evolve further to infinite energy and
compute the $\Lambda$-parameter and the renormalization group invariant
quark masses. Inserted into perturbative expressions these provide
predictions for jet cross sections or other high energy observables. 
In the graph all
arrows correspond to relations in the continuum;
the whole strategy is designed such that lattice calculations
for these relations can be extrapolated to the continuum limit.  

For the practical success of the approach, the finite volume coupling
(as well as the corresponding quark mass) must satisfy a number of criteria.
\begin{itemize}
     \item{They should have an easy perturbative expansion, such that
           the $\beta$-function (and $\tau$-function, which describes the 
           evolution of the running masses) can be computed to 
           sufficient
           order.} 
     \item{They should be easy to calculate in MC (small variance!).}
     \item{Discretization errors must be small to allow
           for safe extrapolations to the continuum limit.}   
\end{itemize}
Careful consideration of the above points led to the introduction
of renormalized coupling and quark mass through the 
Schr\"odinger functional (SF) of QCD 
\myref{alphaII,alphaIII,StefanI,letter}.
We introduce the SF in the following section. 
In the Yang-Mills theory, an alternative 
finite volume coupling was introduced in \cite{alphaTPI} and
studied in detail in \cite{alphaTPII,alphaTPSF}.

The criteria \eq{e_conditions} apply quite generally to any
scale dependent renormalization, e.g. the one described in 
Sect.~\ref{s_sdr}~b. Although the details
of the finite size technique have 
not yet been developed for these cases, the same
strategy can be applied. This will certainly be the subject of future 
research. 
So far, the approach has been to search for a ``window'' where
$q$ is high enough to apply PT but not too close to $a^{-1}$ 
\myref{renor_roma}. An essential advantage of the details of the 
approach of \cite{renor_roma} as applied to the
renormalization of composite quark operators 
is its simplicity: formulating the renormalization conditions
in a {\it MOM}-scheme, 
one may use results from perturbation theory in infinite volume
in the perturbative part of the matching. Since, however,
high energies $q$ can 
not be reached in this approach, we will not discuss it further and refer to
\cite{renor_roma_appl,renor_roma_appl2} for an account of the present 
status and further references, instead. In particular, in the latter 
reference it can be seen, how non-trivial it is to have
a ``window'' where both perturbation theory can be applied 
and lattice artifacts are small.

\paragraph{Note.} \eq{e_conditions} has been written for
the Yang-Mills theory.
In full QCD, finite size effects will be
more important and one should replace $\sqrt{K} \to m_\pi$, resulting
in a more stringent 
requirement. 

\section{The Schr\"odinger functional}

We want to introduce a specific finite volume scheme that
fulfills all the requirements explained in the previous section. 
It is defined from the SF of QCD, which we
introduce below. For simplicity we restrict the discussion to the 
pure gauge theory except for \sect{s_Q} and \sect{s_Rm}.
Apart from the latter subsections,
the presentation follows closely \cite{alphaII}; we refer to this work
for further details as well as proofs of the properties described below. 
 
\begin{figure}[ht]
\centerline{
\psfig{file=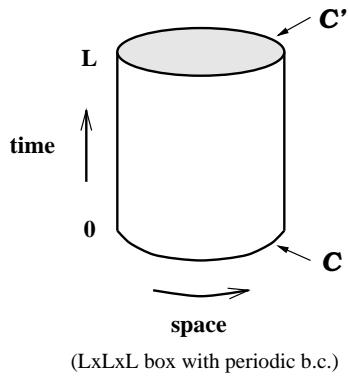,%
width=4.5cm}
}
\vspace{-0.0cm}
\caption{Illustration of the Schr\"odinger functional.\label{f_SF}}
\end{figure}

\subsection{Definition}

Here, we give a formal definition of the SF in the
Yang-Mills theory in continuum space-time,
noting that a rigorous treatment is possible in the lattice regularized 
theory.

Space-time is taken to be a cylinder illustrated in \fig{f_SF}.
We impose  Dirichlet boundary conditions for the vector 
potentials\footnote{We use anti-hermitian vector potentials. E.g.
in the gauge 
group SU(2), we have $A_\mu(x) = A_\mu^a(x) \tau^a/(2i) $,
in terms of the Pauli-matrices  $\tau^a$.
}
in time,
\bes
  A_k(x)&=& \left\{ \begin{array}{ll}
             \bvalue_k^{\Lambda}({\bf x}) & \mbox{at} \quad x_0=0 \\
             \bvalue'_k({\bf x})          & \mbox{at} \quad x_0=L
             \end{array}
          \right.  \enspace ,  \label{e_bcA}
\ees
where $C$, $C'$ are classical gauge potentials and $A^{\Lambda}$
denotes the gauge transform of $A$,
\bes          
  A_k^{\Lambda}({\bf x})&=&
  \Lambda({\bf x})A_k({\bf x})\Lambda({\bf x})^{-1}+
  \Lambda({\bf x})\partial_k\Lambda({\bf x})^{-1}, 
   \qquad \Lambda \in \SU \enspace . \label{e_gaugetrafo}
\ees
In space, we impose periodic boundary conditions,
\bes         
 A_k(x+L\hat k)&=& A_k(x), \qquad   \Lambda({\bf x}+L\hat k)= \Lambda({\bf x}) 
 \enspace .     
\ees
The (Euclidean) partition function with these boundary conditions 
defines the SF,
\bes
  { \schrodinger[\bvalue',\bvalue]} &\equiv&
  \int\rmD[\Lambda]\int\rmD[A]\, 
  \rme^{-\Sg[A]} \enspace,  \label{e_SF}   \label{e_SFdef}\\
  \Sg[A]&=&-{1\over2g_0^2}\int\rmd^4x\,
  \tr\left\{F_{\mu\nu}F_{\mu\nu}\right\}, \nonumber \\ 
  F_{\mu\nu}&=&\partial_{\mu}A_{\nu}-\partial_{\nu}A_{\mu}
  +[A_{\mu},A_{\nu}] \enspace ,\nonumber \\
  \qquad\rmD[A]&=&\prod_{{\bf x},\mu,a}\rmd A_\mu^a({x}),
  \qquad\rmD[\Lambda]=\prod_{\bf x}\rmd\Lambda({\bf x}) \enspace . \nonumber
\ees
Here $\rmd\Lambda({\bf x})$ denotes the Haar measure of $\SU$. 
It is easy to
show that the SF is a gauge invariant functional of the boundary fields,
\bes
  \schrodinger[\bvalue'^{\Omega'},\bvalue^{\Omega}] = 
  \schrodinger[\bvalue',\bvalue] \enspace ,
\ees
where also large gauge transformations are permitted. 
The invariance
under the latter is an automatic property of the SF
defined on a lattice, while in the continuum
formulation it is enforced by the integral over $\Lambda$
in \eq{e_SF}. 

\subsection{Quantum mechanical interpretation}

The SF is the quantum mechanical transition amplitude 
from a state $|\bvalue\rangle$ to a state $|\bvalue'\rangle$
after a (Euclidean) 
time $L$.
To explain the meaning of this statement of the SF,
we introduce the 
Schr\"odinger representation. The Hilbert space consists of 
wave-functionals $\Psi[A]$ which are functionals of the spatial 
components of the vector potentials,
$A_k^a({\bf x})$. 
The canonically conjugate field variables are represented by functional 
derivatives,
$
  E_{k}^a({\bf x})
  ={1\over i}{\delta\over\delta A_k^a({\bf x})}
$, and a scalar product is given by
\bes
  \langle\Psi|\Psi'\rangle=\int\rmD[A]\,\Psi[A]^*\Psi'[A],
  \qquad\rmD[A]=\prod_{{\bf x},k,a}\rmd A_k^a({\bf x}) \enspace .
\ees
The Hamilton operator,   
\bes
  \ham
  =\int_0^L\rmd^3x\,\left\{
  {g_0^2\over2}E_{k}^a({\bf x})E_{k}^a({\bf x})+
  {1\over4g_0^2}F_{kl}^a({\bf x})F_{kl}^a({\bf x})\right\} \enspace ,
\ees
commutes with the
projector, $\projector$, onto the physical subspace of the 
Hilbert space (i.e. the space of gauge invariant states), where 
$\projector$ acts as 
 \bes
  \projector\psi[A]=\int\rmD[\Lambda]\,\psi[A^\Lambda]
 \enspace .
\ees
Finally, each 
classical gauge field defines a state 
$|\bvalue\rangle$ through
\bes
  \langle \bvalue|\Psi\rangle=\Psi[\bvalue] \enspace .
\ees
After these definitions, the 
quantum mechanical representation of the SF is given by
 \bes
  \schrodinger[\bvalue',\bvalue]&=&
  \langle \bvalue'|\rme^{-\ham T}\projector|\bvalue\rangle \nonumber \\
  &=&
  \sum_{n=0}^{\infty} \rme^{-E_nT}
  \Psi_n[\bvalue'] \Psi_n[\bvalue]^* \enspace . \label{e_QM1}
\ees
In the lattice formulation, (\ref{e_QM1}) can be derived 
rigorously and is valid with real energy eigenvalues $E_n$.

\subsection{Background field \label{s_BF}}

A complementary aspect of the SF is that it allows a treatment
of QCD in a color background field in an unambiguous way.
Let us assume that we have a solution $B$ of the equations of motion,
which satisfies also the boundary conditions \eq{e_bcA}.
If, in addition,
\bes
S[A] > S[{B}] \label{e_minim}
\ees
for all gauge fields $A$ that are not equal to a gauge transform $B^\Omega$
of $B$, then we call $B$ the background field 
(induced by the boundary conditions).
Here, $\Omega(x)$ is a gauge transformation defined for all
$x$ in the cylinder and its boundary and $B^\Omega$ is the
corresponding generalization of \eq{e_gaugetrafo}.
Background fields $B$, satisfying these conditions are known; we 
will describe a particular family of fields, later.

Due to \eq{e_minim}, 
fields close to $B$ dominate the path integral 
for weak coupling $g_0$ and the 
effective  action,
\bes
  \effaction[\bfield] &\equiv&
  -\ln \schrodinger\left[\bvalue',\bvalue\right]  \enspace ,\label{e_eff_act}
\ees
has a regular 
perturbative expansion,  
\bes
  \effaction[\bfield] &=&
 {1 \over g_0^{2}}\effaction_0[\bfield]+
  \effaction_1[\bfield]+g_0^2\effaction_2[\bfield]
  +\ldots  \enspace ,\label{e_eff_act_pt} 
  \\
  \effaction_0[\bfield] &\equiv& g_0^2S[\bfield] \enspace . \nonumber
\ees
Above we have used that due to our assumptions, the background field, 
$\bfield$, 
and the boundary values $\bvalue,\bvalue'$ are in one-to-one correspondence
and have taken $\bfield$ as the argument of $\effaction$.

\subsection{Perturbative expansion}

For the construction of the SF-scheme as a renormalization scheme,
one needs to study the renormalization properties of the functional,
$\schrodinger$. 
L\"uscher et al. (1992) have performed a one-loop calculation for
arbitrary background field. The calculation is done
in dimensional regularization with 
 $3-2\varepsilon$ space dimensions and one time dimension. 
One expands the field $A$ in terms of the
background field and a fluctuation field, $q$, as
\bes
  A_{\mu}(x)=\bfield_{\mu}(x)+g_0\qfield_{\mu}(x) \enspace .
\ees
Then one 
adds a gauge fixing term (``background field gauge'') and 
the corresponding Fadeev-Popov term. Of course, care must be taken 
about the proper boundary conditions in all these expressions.
Integration over the quantum field and the ghost fields then
gives
\bes
  \effaction_1[\bfield]=\frac{1}{2}\ln\det\deltaonehat-
  \ln\det\deltazerohat \enspace ,
\ees
where $\deltaonehat$ is the fluctuation operator 
and $\deltazerohat$ the Fadeev-Popov operator. 
The result can be cast in the form
\bes
  \effaction_1[\bfield]
  \mathrel{\mathop=_{\varepsilon\to0}}
  &&-{b_0 \over \varepsilon}
  \,\effaction_0[\bfield]
  +\rmO(1) \enspace , 
\ees
with the important result that the only (for $\varepsilon\to0$) singular term 
is proportional to $\Gamma_0$.  

After renormalization of the coupling, i.e. the
replacement of the bare coupling by $\gbarMSbar$ via 
\bes
     g_0^2 =&& \bar{\mu}^{2\varepsilon} \gbarMSbar^2(\mu) 
         [ 1+ z_1(\varepsilon)\gbarMSbar^2(\mu)],
         \quad z_1(\varepsilon)=-{b_0 \over \varepsilon} \enspace ,
\ees
the effective action is finite,
\bes
   \effaction[\bfield]_{\varepsilon=0}
  &=&\left\{{ 1\over\gbarMSbar^2}-
  b_0
  \left[\ln\mu^2-\frac{1}{16\pi^2}\right]\right\}
  \effaction_0[\bfield]
  \nonumber \\
  &&-\frac{1}{2}\zetaprime{\deltaone}+\zetaprime{\deltazero}
  +\rmO(\gbarMSbar^2) \\
  \zetaprime{\Delta}&=&
  \left.{\rmd\over\rmd s}\zetafunc{s}{\Delta}\right|_{s=0}, \qquad
   \zetafunc{s}{\Delta}=\Tr\,\Delta^{-s} \enspace . \nonumber
\ees
Here, $\zetaprime{\Delta}$ is a complicated functional of $\bfield$,
which is not known analytically but can be
evaluated numerically for specific choices of $\bfield$.

The important result of this calculation is that (apart from
field independent terms that have been dropped everywhere)
the SF is finite after eliminating $g_0$ in favor of $\gbarMSbar$.
The presence of the boundaries does {\it not} introduce any
extra divergences. In the following subsection we argue 
that this property is correct in general, not just in
one-loop approximation.

\subsection{General renormalization properties}

The relevant question here is whether local quantum field
theories formulated on space-time manifolds {\it with boundaries}
develop divergences that are not present in the absence
of boundaries (periodic boundary conditions or infinite space-time).
In general the answer is ``yes, such additional divergences 
exist''.
In particular, Symanzik  studied the $ \phi^4$-theory with 
SF boundary conditions \myref{SymanzikSFI}.
In a proof valid to all orders of perturbation theory 
he
was able to show 
that the SF is finite after
\begin{itemize}
 \item{renormalization of the self-coupling, $\lambda$,  and the
       mass, $m$,}
 \item{{\it and} the addition of the boundary counter-terms
\bes
  \int_{x^0=T}\rmd^3x\,
  \left\{Z_1\phi^2+Z_2\phi\partial_0\phi\right\}+
  \int_{x^0=0}\rmd^3x\,
  \left\{Z_1\phi^2-Z_2\phi\partial_0\phi\right\} \enspace .
\ees
  }
\end{itemize}

In other words, in addition to the standard renormalizations,
one has to add counter-terms formed by local composite fields 
integrated over the boundaries.
One expects that in general, all fields with dimension 
$d\leq3$ have to be taken into account.
Already Symanzik conjectured that counter-terms with this property are
sufficient to renormalize the SF of any quantum field theory
in four dimensions. 

Since this conjecture forms
the basis for many applications of the  SF to the study of
renormalization, we note a few points concerning its status.
\begin{itemize}
 \item{As mentioned, a proof to all orders of perturbation theory 
       exists for the $\phi^4$ 
       theory, only.}
 \item{There is no gauge invariant local field with $d\leq3$ in the  Yang--Mills 
       theory. Consequently no additional
       counter-term is necessary in accordance with the 1-loop result described in 
       the previous
       subsection.}
 \item{In the Yang--Mills theory it has been checked also by
       explicit 2--loop calculations~\myref{twoloop1,twoloop2}. 
       Numerical, non-perturbative, MC simulations~\myref{alphaIII,alphaTPSF} 
       give further support for its validity.
       }
 \item{It has been shown to be valid in QCD with quarks to 1-loop~\myref{StefanI}. }      
 \item{A straight forward application of power counting in momentum space 
       in order to prove the 
       conjecture is not
       possible due to the missing translation invariance. }
 \end{itemize}
Although a general proof is missing, there is little doubt that Symanzik's 
conjecture is valid in general. Concerning QCD, 
this puts us into the position 
to give an elegant definition of
a renormalized coupling in finite volume.

\subsection{Renormalized coupling}

For the definition of a running coupling we need a quantity which
depends only on one scale. We choose $L B$ such that it
depends only on one dimensionless variable $\bfieldparm$. 
In other words, the strength of the field is scaled as $1/L$.
The background field is assumed to fulfill the requirements of
\sect{s_BF}.
Then, following the above discussion, the derivative
\bes
  \effaction'[\bfield]=
  {\partial\over\partial\bfieldparm}
  \effaction[\bfield] \enspace ,
\ees
is finite when it is expressed in terms of a renormalized coupling
like $\gbarMSbar$ but $\effaction'$ is defined 
non-perturbatively.
From \eq{e_eff_act_pt} we read off immediately that a properly
normalized coupling is given by
\bes 
 \gbar^2(L)=\effaction'_0[\bfield]\bigm/\effaction'[\bfield] \enspace .
  \label{e_gbarsf}
\ees
Since there is only one length scale $L$, it is evident that $\gbar$
defined in this way runs with $L$.

A specific choice for the gauge group $\SUthree$ is
the abelian background field induced by the boundary values~\myref{alphaIII}
\bes
  C^{}_k = \frac{i}{L} \left(
          \begin{array}{ccc} 
               \phi^{}_1 & 0          & 0        \\
                0        & \phi^{}_2  & 0        \\      
                0        & 0          & \phi^{}_3 
         \end{array} \right) \, ,\quad
  C'_k = \frac{i}{L}
         \left( \begin{array}{ccc}
               \phi'_1   & 0          & 0        \\
                0        & \phi'_2    & 0        \\      
                0        & 0          & \phi'_3 
         \end{array} \right) \, ,
 \quad k=1,2,3, 
 \label{e_abelian}       
\ees                      
with
\bes
  \begin{array}{lll}
  \phi^{}_1  = \eta-\frac{\pi}{3},      
    &\quad& \phi'_1 = -\phi^{}_1-\frac{4\pi}{3},  \\[1ex]
  \phi^{}_2  = -\frac12 \eta,       
    &\quad& \phi'_2 = -\phi^{}_3+\frac{2\pi}{3},   \\[1ex]
  \phi^{}_3  = -\frac12\eta+\frac{\pi}{3}, 
    &\quad&\phi'_3  = -\phi^{}_2+\frac{2\pi}{3}.   
    \end{array}
     \label{e_bflds}
\ees
In this case, the derivatives with respect to 
$\eta$ are to be evaluated at $\eta=0$. The 
associated background field,
 \begin{equation}
  B_0=0,\qquad B_k=\left[x_0 C_k' + (L-x_0) C_k^{}\right]/L,
  \quad k=1,2,3 \enspace ,\label{e_BF}
\end{equation}
has a field tensor with  non-vanishing components
\bes
 G_{0k}=\partial_0 B_k=(C_k'-C_k)/L,\quad k=1,2,3 \enspace .\label{e_G0k}
\ees
It is a constant color-electric field. 

\subsection{Quarks \label{s_Q}}

In the end, the real interest is in the renormalization of
QCD and we need to consider the SF with  quarks.
It  
has been discussed in \cite{StefanI}.

Special care has to be taken in formulating the 
Dirichlet boundary conditions for the quark fields; since the Dirac operator
is a first order differential operator, the Dirac equation has a unique 
solution when one half of the components of the fermion fields are
specified on the boundaries. Indeed, a detailed investigation shows
that the boundary condition
\bes
        P_+\psi|_{x_0=0} &=& \rho, \quad P_- \psi|_{x_0=L} = \rho'\, ,
          \qquad P_\pm = \frac{1}{2}(1\pm\gamma_0) \, ,\\
       \psibar P_-|_{x_0=0} &=& \rhobar, 
        \quad \psibar P_+|_{x_0=L} = \rhobar' \, ,
\ees
lead to a quantum mechanical 
interpretation analogous to \eq{e_QM1}. The SF 
\bes
  {\cal Z}[C',\rhobarprime,\rhoprime; C,\rhobar,\rho]=
  \int\rmD[A]\rmD[\,\psi\,]\rmD[\,\psibar\,]\,\rme^{-S[A,\psibar,\psi\,]}
  \label{e_sfqcd}
\ees 
involves an integration over all fields with the specified boundary
values.
The full action may be written as
\bes
S[A,\psibar,\psi\,]&=&\Sg[\psibar,\psi\,]+\Sf[A,\psibar,\psi\,] \nonumber \\ 
\Sf &=& \int \rmd^4 x \, \psibar(x)  [\gamma_\mu D_\mu +m]  \psi(x) 
    \label{e_fermact} \\
    && - \int \rmd^3 {\bf x} \,[ \psibar(x) P_- \psi(x)]_{x_0=0}
       - \int \rmd^3 {\bf x} \,[ \psibar(x) P_+ \psi(x)]_{x_0=L}
       \enspace , \nonumber
\ees
with  $\Sg$ as given in \eq{e_SFdef}. In \eq{e_fermact} we use standard 
Euclidean $\gamma$-matrices. The covariant derivative, $D_\mu$, acts as
$D_\mu \psi(x) = \partial_\mu \psi(x) + A_\mu(x) \psi(x)$.

Let us now discuss the renormalization of the SF with quarks.
In contrast to the pure Yang-Mills theory, gauge invariant 
composite fields of dimension three are present in QCD.
Taking into account the boundary conditions 
one finds \myref{StefanI} that the counter-terms,
\bes
    \psibar P_- \psi|_{x_0=0}\, \, {\rm and} \, \, \psibar P_+ \psi|_{x_0=L} 
    \enspace ,
\ees
have to be added to the action with 
weight $1-Z_{\rm b}$ to obtain a finite
renormalized functional. These counter-terms 
are equivalent to a multiplicative renormalization of the
boundary values, 
\bes
  \rho_{\rm R} = Z_{\rm b}^{-1/2} \rho \, , \,\,\ldots \, \,\, , \,
  \rhobar'_{\rm R} = Z_{\rm b}^{-1/2} \rhobar' \enspace .
  \label{Zb}
\ees
It follows that -- apart from 
the renormalization of the coupling and the quark mass -- no
additional renormalization of the SF is necessary for {\it vanishing}
boundary values
$\rho, \ldots,\rhobar'$.
So, after imposing homogeneous boundary conditions for the fermion fields,
a renormalized coupling may be defined as in the previous subsection.

As an important aside,  we point out that the boundary conditions for the 
fermions
introduce a gap into the spectrum of the Dirac operator (at least for
weak couplings). One may hence simulate the lattice
SF for vanishing physical quark 
masses. It is then convenient to supplement the definition of the renormalized
coupling by the requirement $m=0$. In this way, one defines a mass-independent
renormalization scheme with simple renormalization group equations.
In particular, the $\beta$-function remains independent of the quark
mass.

\subsubsection{ Correlation functions}
are given in terms of the expectation values of
any product $\cal O$ of fields,
\bes
  \langle{\cal O}\rangle=\left\{{1\over{\cal Z}}
  \int\rmD[A]\rmD[\,\psi\,]\rmD[\,\psibar\,]\,{\cal O}\,
  \rme^{-S[A,\psibar,\psi\,]}\right\}_
  {\rhobarprime=\rhoprime=\rhobar=\rho=0} \enspace,
\ees
evaluated for vanishing boundary values
$\rho, \ldots,\rhobar'$. Apart from the gauge field and the quark and 
anti-quark fields integrated over, $\cal O$ may involve the 
``boundary fields"~\myref{paper1}
\bes
  \zeta({\bf x})&=&{\delta\over\delta\rhobar({\bf x})},
  \qquad
  \zetabar({\bf x})=-{\delta\over\delta\rho({\bf x})},
  \nonumber \\
  \zeta'({\bf x})&=&{\delta\over\delta\rhobarprime({\bf x})},
  \qquad
  \zetabarprime({\bf x})=-{\delta\over\delta\rhoprime({\bf x})} \enspace .
\ees
An application of fermionic correlation functions 
including the boundary fields
is the definition of the renormalized quark mass in the
SF scheme to be discussed next.

\subsection{Renormalized mass \label{s_Rm}}

Just as in the case of the coupling constant, there is a great freedom in 
defining renormalized quark masses. A natural starting point 
is the PCAC relation which expresses the divergence of the axial current
\eq{e_currents} in terms of the associated pseudo-scalar density,
\bes            
  P^a(x) &=& 
            \psibar(x)\dirac{5}\frac{1}{2}\tau^a\psi(x) \enspace ,
  \label{e_density}          
\ees
via
\bes
 \partial_\mu A_{\mu}^a(x) = 2 m P^a(x)\enspace .
 \label{e_PCAC}
\ees
This operator identity is easily derived at the classical level 
(cf. \sect{s_curr}).
After renormalizing the operators,
\bes
 (\ar)_{\mu}^a &=& \za A_{\mu}^a \enspace ,\nonumber \\
  \pr^a &=& \zp P^a \enspace , \label{e_zazp}
\ees
a renormalized current quark mass may be defined by
\bes
 \mbar = {\za \over \zp } m \enspace . 
 \label{e_mbar}
\ees
Here, $m$, is to be taken from \eq{e_PCAC} inserted into an arbitrary 
correlation function and $\za$ can be determined unambiguously 
as mentioned in \sect{s_finiter}. Note that $m$ does not
depend on which correlation function is used because the PCAC relation is an 
operator identity.
The
definition of $\mbar$ is completed by supplementing
\eq{e_zazp} with a specific normalization
condition for the pseudo-scalar density. $\mbar$ then inherits its scheme-
and scale-dependence from the corresponding dependence of
$\pr$. Such a normalization condition may be imposed through infinite
volume correlation functions. Since we want to be  able to compute the
running mass for large energy scales, we do, however, need a finite volume
definition. This is readily given in terms of
correlation functions in the SF.
\begin{figure}[ht]
\vspace{ -0.0cm}
\centerline{
\psfig{file=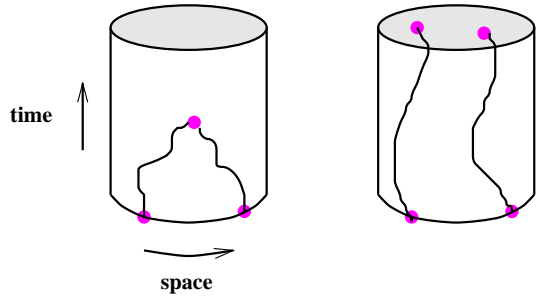,%
width=7cm}}
\vspace{-0.0cm}
\caption{$\fp$ (left) and $f_1$ (right) in terms of quark propagators.
\label{f_matrixelements}}
\end{figure}

To start with, let us define (isovector) pseudo-scalar fields at the 
boundary of the SF,
\bes
  \op{}^a &=& \int \rmd^3 {\bf u} \int \rmd^3 {\bf v}
  \,\,\zetabar({\bf u})\dirac{5}\frac{1}{2}
  \tau^a\zeta({\bf v}), 
  \nonumber \\
  \opprime{}^a &=& \int \rmd^3 {\bf u} \int \rmd^3 {\bf v}
  \,\,\zetabarprime({\bf u})\dirac{5}\frac{1}{2}
  \tau^a\zeta'({\bf v}) 
  \enspace,  \label{e_boundops}
\ees
to build up the correlation functions 
\bes
 \fp(x_0) &=& - \frac13 \langle P^a(x)  \op{}^a\rangle \enspace ,\nonumber \\
 f_1 &=& \langle\opprime{}^a\op{}^a\rangle \enspace ,
\ees
which are illustrated in \fig{f_matrixelements}.

We then form the ratio 
\bes
\zp = {\rm const.} \sqrt{f_1} / \fp(x)|_{x_0=L/2}  
\enspace , \label{e_zp}
\ees
such that the renormalization of the boundary quark fields, \eq{Zb}, 
cancels out. 
The proportionality
constant is to be chosen such that $\zp=1$ at tree level. To define the scheme
completely one needs to further specify the boundary values $C,C'$ and the
boundary conditions for the quark fields in space. These details 
are of no importance, here. 

We rather mention some more basic points about this
renormalization scheme. 
Just like in the case of the
running coupling, the only physical scale that exists in our definitions 
\eq{e_mbar},\eq{e_zp} is the linear dimension of the SF, the length scale, $L$.
So the mass $\mbar(L)$ runs with $L$.
We have already emphasized
that $\gbar$ is to be evaluated at zero quark mass. It is
advantageous to do the same for $\zp$. In this way
we define a mass-independent renormalization scheme, with simple
renormalization
group equations.

By construction, the SF scheme is non-perturbative
and independent of a specific regularization. For a concrete non-perturbative
computation, we do, however, need to evaluate the expectation values
by a MC-simulation of the corresponding lattice theory. We proceed
to introduce the lattice formulation of the SF.

\subsection{Lattice formulation}
A detailed 
knowledge of the form of the lattice action is 
not required for an understanding of the following sections. 
Nevertheless, we 
give a definition of the SF in lattice regularization. This is done
both for completeness and
because it allows us to obtain a first impression about the
size of discretization errors.

We choose a hyper-cubic Euclidean lattice with spacing $a$.
A gauge field $U$ on the lattice is an assignment of a matrix 
$U(x,\mu)\in\SUn$ to every lattice point $x$ and direction 
$\mu=0,1,2,3$.
Quark and anti-quark fields, $\psi(x)$ and $\psibar(x)$,
reside on the lattice sites and  
carry Dirac, color and flavor indices as in the continuum.
To be able to write the quark action in
an elegant form it is useful to 
extend the fields, initially defined only inside the SF manifold (cf. \fig{f_SF})
to all times $x_0$ by ``padding" with zeros.
In the case of the quark field one sets
$$
  \psi(x)=0\quad\hbox{if $x_0<0$ or $x_0>L$},
$$
and
$$
  P_{-}\psi(x)|_{x_0=0}=
  P_{+}\psi(x)|_{x_0=L}=0,
$$
and similarly for the anti-quark field.
Gauge field variables that reside outside the manifold are set to 1.

We may then write the fermionic action as a sum
over all space-time points without restrictions for the time-coordinate,
\bes 
  \Sf[U,\bar\psi,\psi] &=& a^4\sum_{x}\bar\psi(D+m_0)\psi , \label{e_Sf}
\ees
and with the
standard Wilson-Dirac operator,
\begin{equation}
  D = \frac12 \sum_{\mu=0}^3
  \{\gamma_\mu(\nabla_\mu^\ast+\nabla_\mu^{})- a\nabla_\mu^\ast\nabla_\mu^{}\} \, .
\label{e_Dlat}
\end{equation}
Here, forward and backward covariant derivatives, 
\begin{eqnarray}
  \nabla_\mu^{}\psi(x)     &=& \frac1a [U(x,\mu)\psi(x+a\hat\mu)-\psi(x)],\\
  \nabla_\mu^{\ast}\psi(x) &=&
   \frac1a [\psi(x)-U(x-a\hat\mu,\mu)^{-1}\psi(x-a\hat\mu)] \enspace ,
\end{eqnarray}
are used and $m_0$ is to be understood as a diagonal matrix in flavor space
with elements $m_0^f$.

The gauge field action $\Sg$ is a sum over all oriented 
plaquettes $p$ on the lattice, with the weight factors $w(p)$, and 
the parallel transporters $U(p)$  around $p$,  
\bes
 \Sg[U]={1\over g_0^2}\sum_p w(p)\,\tr\{1-U(p)\} \enspace.
 \ees
The weights 
$w(p)$ are 1 for plaquettes in the interior and 
\begin{equation}
 w(p)=
 \left\{ 
 \begin{array}{ll}
         {\scriptstyle \frac12} c_s &
      \mbox{if $p$ is a spatial plaquette at $x_0=0$ or $x_0=L$},\\
      c_t &
     \mbox{if $p$ is time-like and attached to a boundary plane.} 
   \end{array}
   \right.
\end{equation} 
The choice $c_s=c_t=1$ corresponds to the standard Wilson action.
However, these parameters can be tuned in order to reduce lattice artifacts,
as will be briefly discussed below.

With these ingredients, the  
path integral representation of the Schr\"odinger functional 
reads~\myref{StefanI},
\bes
  {\cal Z} &=& \int{\rm D}[\psi]{\rm D}[\bar\psi\,]{\rm D}[U]\,
                 {\rm e}^{-S}\,, \quad S=\Sf+\Sg \enspace , \\
           &&{\rm D}[U]  = \prod _{x,\mu} \rmd U(x,\mu) \enspace , \nonumber  
\ees
with the Haar measure $\rmd U$.

\subsubsection{Boundary conditions and the background field.}

The boundary conditions for the lattice gauge fields 
may be obtained from the continuum boundary values by forming the 
appropriate parallel transporters from $x+a\hat{k}$ to $x$ 
at $x_0=0$ and $x_0=L$. 
For the constant abelian boundary
fields $C$ and $C'$
that we considered before, they are simply
\begin{equation}
  U(x,k)|_{x_0=0}=\exp(aC_k^{}),\qquad U(x,k)|_{x_0=L}=\exp(aC'_k),
  \label{e_latbc}
\end{equation}
for $k=1,2,3$.
All other boundary conditions are as in the continuum.

For the case of \eq{e_abelian},\eq{e_bflds}, 
the boundary conditions~(\ref{e_latbc}) lead to 
a unique (up to gauge transformations) minimal action configuration $V$,
the lattice background field. It can be expressed 
in terms of $B$~(\ref{e_BF}), 
\begin{equation}
  V(x,\mu)=\exp\left\{a B_\mu(x)\right\} \enspace .
\end{equation}

\subsubsection{Lattice artifacts.}

Now we want to get a first impression about the dependence of the lattice
SF on the value of the lattice spacing. In other words we study
lattice artifacts.
At lowest order in the bare coupling we have, just like in the 
continuum,
\bes
  \effaction = 
 {1 \over g_0^{2}}\effaction_0[\bfieldlat]+ \rmO((g_0)^0)
  \, , \quad
  \effaction_0[\bfieldlat] \equiv g_0^2 \Sg[\bfieldlat] \enspace . 
  \label{e_eff_act_pt_lat}
\ees
Furthermore one easily finds the action for small lattice spacings, 
\bes
  \Sg[\bfieldlat]&=&\left[1+(1-\ct) \frac{2a}{L}\right] 
  {3 L^{4}\over g_0^2}
  \sum_{\alpha=1}^N
  \left\{{2\over a^2}\sin\left[{a^2\over2L^2}
  \left(\phi'_{\alpha}-\phi_{\alpha}\right)
  \right]\right\}^2 \nonumber \\
  &=& 
 {3 \over g_0^2}
  \sum_{\alpha=1}^N
  \left(\phi'_{\alpha}-\phi_{\alpha}\right)^2 
  \left[1+( 1-\ct) { \frac{2a}{L} +\rmO(a^4)}\right] \enspace .
  \label{e_act_expanded}
\ees
We observe: at tree-level of 
perturbation theory, all linear lattice artifacts
are removed when one sets $\ct=1$. 
Beyond tree-level, one has to 
tune the coefficient $\ct$ as a function
of the bare coupling. We will show the effect, when this is done to first
order in $g_0^2$, below. Note that the existence of linear $\rmO(a)$ 
errors in the Yang-Mills theory is special to the SF; they originate
from dimension four operators $F_{0k}F_{0k}$ and $F_{kl}F_{kl}$
which are irrelevant terms (i.e. they carry an explicit factor of
the lattice spacing) when they are integrated over the surfaces.
$\cs$, which can be tuned to cancel the effects of $F_{kl}F_{kl}$,
does not appear for the electric field that we discussed above.

Once quark fields are present, there are more irrelevant operators that can 
generate $\rmO(a)$ effects as discussed in detail in \cite{paper1}. 
Here we emphasize a different feature of \eq{e_act_expanded}:
once the $\rmO(a)$-terms are canceled, the remaining $a$-effects are tiny.
This special feature of the abelian background field is most welcome
for the numerical computation of the running coupling; it allows for
reliable extrapolations to the continuum limit.

\subsubsection{Explicit expression for $\effaction'$.}

Let us finally explain that $\effaction'$ is an observable that can easily
be calculated in a MC simulation. From its definition we find immediately
\bes
{\Gamma'} &=& -{\partial \over \partial\eta}
 \ln  \left\{ \int{ {\rm D}[\psi]{\rm D}[\bar\psi\,]{\rm D}[U]\,
                 {\rm e}^{-S} } \right\} =
                \left\langle {\partial S \over \partial\eta}
                                           \right\rangle  \enspace .
\ees
The derivative $\frac{\partial S}{\partial\eta}$ evaluates
to the (color 8 component of the) electric field
at the boundary,
\bes
{\partial S \over \partial\eta}&=&
 - {2 \over g_0^2 \, L} a^3 \sum_{\bf x}\left\{E_k^8({\bf x}) - 
                (E_k^8)'({\bf x}) \right\} \enspace , \\
 E_k^8({\bf x})&=&  {1 \over a^2}\Re \tr \left\{ i \lambda_8 U(x,k) U(x+a\hat k,0) 
                  U(x+a\hat 0,k)^{-1} U(x,0)^{-1}
               \right\}_{x_0=0} \enspace ,\nonumber
\ees
where $\lambda_8={\rm diag}(1,-1/2,-1/2)$.
(A similar expression holds for $( E_k^8)'({\bf x}) $).
The renormalized coupling is therefore given in terms of the expectation value
of a local operator; no correlation function is involved. 
This means that it is easy and fast in computer time
to evaluate it. It further turns out that a 
good statistical precision is reached with a moderate size statistical
ensemble.

\section{The computation of $\alpha(q)$}

We are now in the position to explain the details of \fig{f_strategy}
\myref{alphaI,alphaIII,lat97}. The problem has been solved in the 
SU(3) Yang-Mills theory. In the present
context, this is of course equivalent to
the quenched approximation of QCD or the limit of zero flavors.
We will therefore also refer to results in quenched QCD.

Our central observable is the step scaling function that describes
the scale-evolution of the coupling, i.e. moving vertically in
\fig{f_strategy}. The analogous function for
the running quark mass will be discussed in the following section.

\subsection{The step scaling function}
\begin{figure}
\hspace{0cm}
\vspace{-1.4cm}

\centerline{
\psfig{file=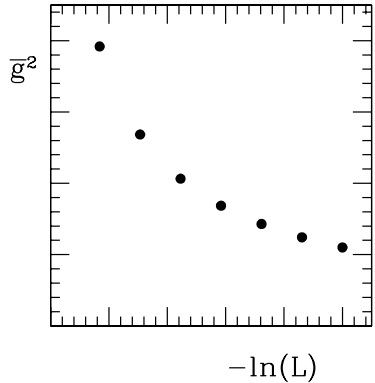,width=7cm}
}
\vspace{-0.5cm}
\caption{Schematic plot of the running coupling constructed 
from the step scaling function $\sigma$.
\label{f_ssf}}
\end{figure}

We start from a given value of the coupling, $u=\gbar^2(L)$.
When we change the length scale by a factor $s$, the coupling
has a value $\gbar^2(sL)=u'$ .
The step scaling function, $\sigma$ 
is then defined as
\bes
 \sigma(s,u)&=&u' \enspace .
\ees
The interpretation is obvious. $\sigma(s,u)$ is a 
discrete $\beta$-function. Its knowledge allows for the 
recursive construction
of the running coupling at discrete values of the
length scale,
\bes
  u_k = \gbar^2(s^{-k} L) \enspace , \label{e_uk}
\ees
once a starting value $u_0 = \gbar^2(L)$ is specified
(cf. \fig{f_ssf}).
$\sigma$, which is readily expressed as an integral 
of the $\beta$-function, has a perturbative expansion 
\bes
 \sigma(s,u)&=& u + 2 b_0 \ln(s) u^2 +\ldots  \enspace .
\ees

On a lattice with finite spacing, $a$, the step scaling function 
will have an additional dependence on 
the resolution $a/L$. We define
\bes
 \Sigma(s,u,a/L)&=&u'\enspace ,
\ees
with   
\bes
 \gbar^2(L)&=&u, \quad  \gbar^2(sL)=u'\, , \quad 
 \mbox{$g_0$ fixed, $L/a$ fixed} \enspace .
\ees
The continuum limit $\sigma(s,u) = \Sigma(s,u,0)$ is then 
reached by performing calculations for several
different resolutions and extrapolation $a/L \to 0$. 
In detail, one performs the following steps:
\begin{itemize}
 \item[1.] Choose a lattice with $L/a$ points in each direction.
 \item[2.] Tune the bare coupling $g_0$ such that the renormalized
           coupling $\gbar^2(L)$ has the value $u$.
 \item[3.] At the same value of $g_0$, simulate a lattice 
           with twice the linear size; compute 
           $u'=\gbar^2(2L)$. This determines the lattice step scaling function
           $\Sigma(2,u,a/L)$.
 \item[4.] Repeat steps 1.--3. with different resolutions $L/a$ and extrapolate
           $a/L \to 0$.       
\end{itemize}
Note that step 2. takes care of the renormalization and 3. determines
the evolution of the {\it renormalized} coupling.

\begin{figure}
\vspace{-2.6cm}

\psfig{file=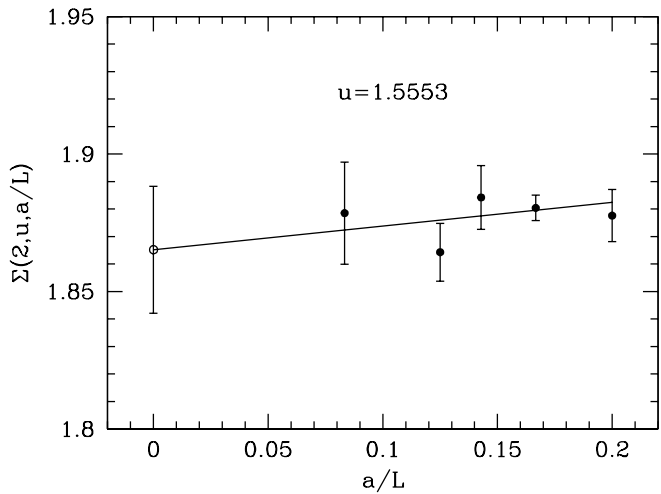,width=10cm}
\vspace{-1.0cm}
\caption{Typical example for the lattice step scaling function
         after 1-loop improvement. The continuum
         limit (circle) is reached by linear extrapolation.
       \label{f_Sigma_2.1} }
\end{figure}

Sample numerical results are displayed in \fig{f_Sigma_2.1}. 
The coupling used is exactly the one defined in the previous section
and the calculation is done in the theory without fermions.
One observes that the dependence on the resolution is very weak, 
in fact it is not observable with the precision of the data in 
\fig{f_Sigma_2.1}. We now investigate in more detail how the
continuum limit  of $\Sigma$ is reached.  As a first step,
we turn to
perturbation theory.

\subsection{Lattice spacing effects in perturbation theory}

Symanzik has investigated the cutoff dependence of field theories
in perturbation theory~\myref{Symanzik}. Generalizing
his discussion to the present
case, one concludes that the lattice spacing effects
have the expansion
\bes
 {\Sigma(2,u,a/L)-\sigma(2,u) \over \sigma(2,u)} &=& \delta_1(a/L) \, u +
 \delta_2(a/L) \, u^2+ \ldots 
  \label{e_delta} \\ 
 \delta_n(a/L)  &{\buildrel {a/L}\rightarrow0\over\sim } &
 \sum_{k=0}^n e_{k,n} [\ln(\frac aL)]^k { \left(\frac aL\right)} +
              d_{k,n} [\ln(\frac aL)]^k { \left(\frac aL\right)^2} + \ldots
              \enspace .
 \nonumber
\ees
We expect that the continuum limit 
is reached with corrections $\rmO(a/L)$ also beyond perturbation
theory. In this context $\rmO(a/L)$
summarizes terms that contain at least one power of
$a/L$ and may be modified by logarithmic corrections as it is the case
in \eq{e_delta}.
To motivate this expectation recall \sect{s_Io}, where we explained
that lattice artifacts correspond to irrelevant
operators\footnote{For a more precise meaning of this terms
one must discuss Symanzik's effective theory. We refer the reader 
to \cite{paper1} for such a discussion.}, 
which carry explicit factors of the lattice spacing. Of course,
an additional $a$-dependence comes from their anomalous dimension,
but in an asymptotically free theory such as QCD, this just corresponds 
to a logarithmic (in $a$) modification. 

\begin{figure}[ht]
\vspace{-2.8cm}

\centerline{
\psfig{file=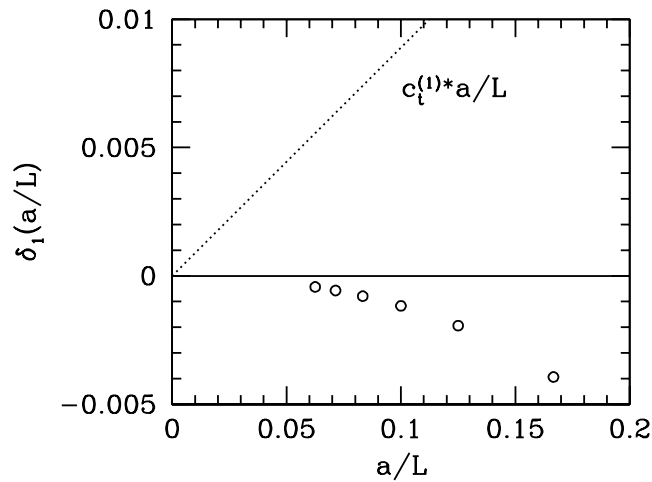,width=10.5cm}
}
\vspace{-1.5cm}
\caption{Lattice artifacts at 1-loop order. The circles show
         $\delta_1(a/L) $ 
         for the SU(3) Yang-Mills theory with 1-loop improvement.
         The dotted line corresponds to the linear piece in $a$,
         when only tree-level improvement is used, instead.
         \label{f_delta1}}
\end{figure}

As mentioned in the previous section, the lattice artifacts may be
reduced to $\rmO((a/L)^2)$ by canceling the 
leading irrelevant operators. In the case at hand, this is achieved by  a 
proper choice of $\ct(g_0)$.
It is interesting to note, that by using the perturbative
approximation 
\bes
 \ct(g_0) = 1+ \ct^{(1)}g_0^2
\ees
one does not only eliminate $e_{1,n}$ for $ n=0,1$ but also the logarithmic 
terms generated at higher orders are reduced, 
\bes
e_{n,n}=0, \quad e_{n-1,n}=0\enspace .
\ees
For tree-level improvement, $\ct(g_0) = 1$, the corresponding
statement is $e_{n,n}=0$. Heuristically, the latter is easy to understand.
Tree-level improvement means that the propagators and vertices
agree with the continuum ones up to corrections
of order $\rmO(a^2)$. Terms proportional to $a$ can then arise 
only through a linear divergence of the Feynman diagrams. Once this happens, 
one cannot have the maximum number of 
logarithmic divergences any more; consequently 
$e_{n,n}$ vanishes.

To demonstrate further that the abelian field introduced in the previous 
section induces small lattice artifacts, we show $\delta_1(a/L)$ for
the one loop improved case. The term that is canceled by the proper choice 
$\ct^{(1)}=-0.089$ is shown as a dashed line. 
The left over 
$\rmO((a/L)^2)$-terms are  below the 1\% level
for couplings $u\leq2$ and lattice sizes $L/a\geq 6$. 
We now understand better
why the $a/L$-dependence is so small in \fig{f_Sigma_2.1}.

From the investigation of lattice spacing effects in perturbation theory
one expects that one may safely extrapolate to the continuum limit by
a fit
\bes
\Sigma(2,u,a/L) = \sigma(2,u) +{\rm const. } \times a/L \enspace ,
\label{e_extrap}
\ees 
once one has data with a weak dependence on $a/L$, like the ones
in \fig{f_Sigma_2.1}. Such an extrapolation is 
 shown in the figure.

\subsection{The continuum limit -- universality}

\begin{figure}[h]
\vspace{-1.8cm}

\centerline{
\psfig{file=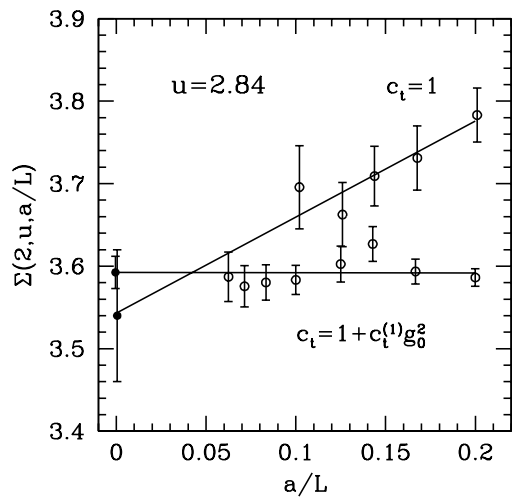,%
width=9.0cm}}
\vspace{-0.8cm}
\caption{Universality test in the SU(2) Yang Mills theory.
         \label{f_universal}}
\end{figure}
Before proceeding with the extraction of the running coupling,
we present some further examples of numerical investigations of the
approach to the continuum 
limit -- and its very existence~\myref{alphaIII,alphaTPSF}. 
The first example is the step scaling function in the 
SU(2) Yang-Mills theory  \myref{alphaTPSF}. Here we can compare 
the step scaling function obtained with two different 
lattice actions, one using tree-level $\Oa$ improvement
and the other one using $\ct$ at 1-loop order.
(\fig{f_universal}).

Not only does one observe a substantial reduction of the $\Oa$-errors
through perturbative improvement, but the very agreement of the two 
calculations when extrapolated to $a=0$, leaves little doubt
that the continuum limit of the SF exists and is independent 
of the lattice action. In turn this also supports the statement
that the SF is renormalized after the renormalization of the coupling 
constant.
 
\begin{figure}
\hspace{-1.5cm}

\psfig{file=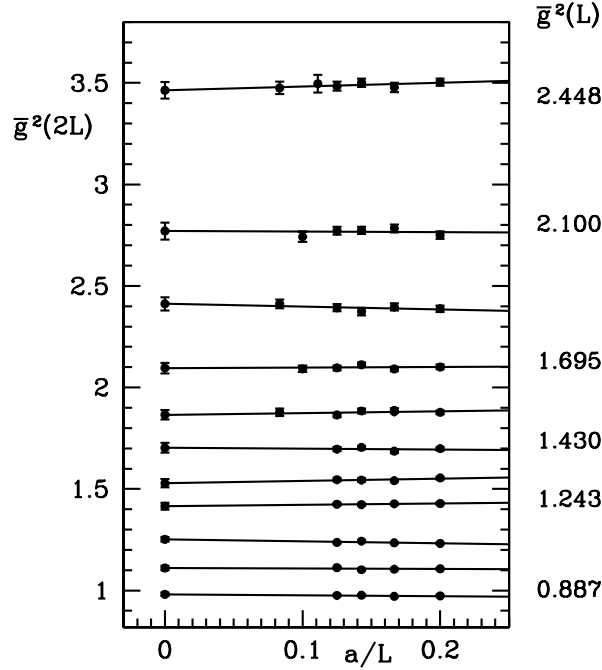,%
width=12cm}
\vspace{-2.3cm}
\caption{Continuum extrapolation of $\sigma(2,u)$ in the SU(3) Yang-Mills
         theory.
         \label{f_sigma}}
\end{figure}

Turning attention back to the gauge group SU(3), we show the calculation
of $\sigma(2,u)$ for a whole series of couplings $u$ in \fig{f_sigma}.

\subsection{The running of the coupling}

\begin{figure}
\vspace{-2.2cm}

\psfig{file=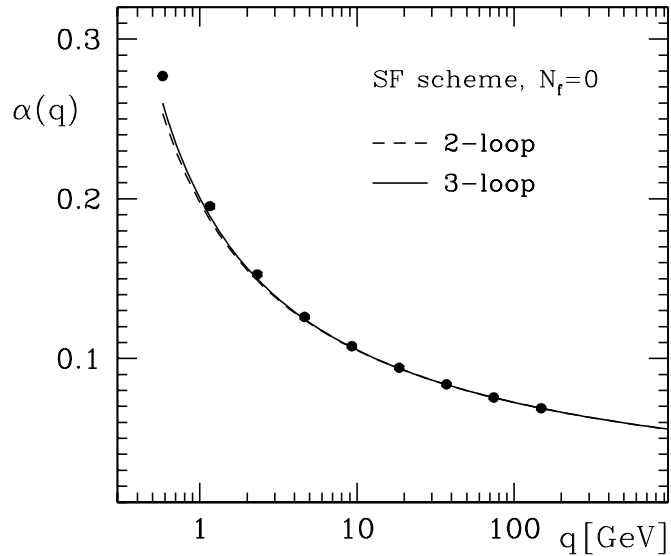,%
width=12cm}
\vspace{-2.0cm}
\caption{The running coupling in  SU(3) Yang-Mills
theory.  Uncertainties are smaller than the size of the symbols.
\label{f_running}}
\end{figure}
We may now use the continuum step scaling function to compute a series 
of couplings \eq{e_uk}. We start at the largest value of the
coupling that was covered by the calculation: $\gbar^2=3.48$. This
defines the largest value of the box size, $\Lmax$, 
\bes
 \gbar^2(\Lmax) = 3.48 \enspace .
\ees
The series of couplings is then obtained for 
$L_k=2^{-k}\Lmax, \, k=0,1,\ldots 8$. It
is shown in \fig{f_running} translated to
$\alpha(q) = \gbar^2(L)/(4\pi), \,\, q=1/L$ 
(We will explain below, 
how one arrives at a $\GeV$-scale in this plot).
The range of couplings 
shown in the figure is the range covered in the non-perturbative 
calculation of the step scaling function. Thus no approximations are 
involved. For comparison, the perturbative evolution is shown starting
at the smallest value of $\alpha$ that was reached.
To be precise, 2-loop accuracy here means that we truncate the 
$\beta$-function at 2 loops and integrate the resulting renormalization
group equation exactly. 
Thanks to the recent work \myref{twoloop2,twoloop3}, we can also compare
to the 3-loop evolution of the coupling. 

It is surprising that the perturbative evolution 
is so precise down to very low energy scales. This property may 
of course not be generalized to other schemes, in particular not
to the $\MSbar$-scheme, where the $\beta$-function is only defined in 
perturbation theory, anyhow.

\subsection{The low energy scale} 
In order to have the coupling as a function of the energy scale
in physical units, we need to know $\Lmax$ in $\fm$, the first horizontal 
relation in \fig{f_strategy}. In QCD, this should be done by computing, 
for example, the product $m_{\rm p} \Lmax$ with $m_{\rm p}$ the proton mass
and then inserting the experimentally  determined value of the proton 
mass. 

At present, results like the ones shown in \fig{f_running}
are available for the Yang-Mills theory, only. 
Therefore, strictly speaking, there is no experimental observable
to take over the role of the  proton 
mass. As a purely theoretical exercise, one could replace the
proton mass by a glueball mass; here,  we choose a length scale, $r_0$,
derived from the force between static quarks, instead~\myref{r0paper}. This quantity can be 
computed with better precision. Also one may argue that the
static force is less influenced by whether one has
dynamical quark loops in the theory or not.

\begin{figure}[ht]
\hspace{1.5cm}
\vspace{-1.7cm}

\centerline{
\psfig{file=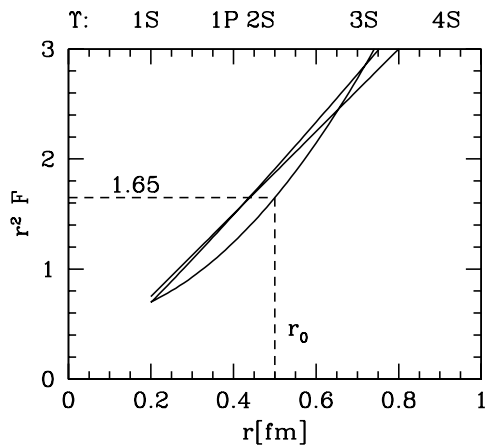,width=9cm}}
\vspace{-1.0cm}
\protect
        \caption{The dimensionless combination $r^2F(r)$. The different curves
         show phenomenologically successful potential 
         models~\protect\myref{Cornell,Martin,Quigg-Rosner}. 
         The labels on the 
         top of
         the graph give the approximate values of the r.m.s-radii of
         the bound states.
         \label{f_def_r0}}
\end{figure}
On the theoretical level, $r_0$, has a precise definition. 
One evaluates the force $F(r)$ between an external, static, 
quark--anti-quark pair as a function of the distance $r$. 
The radius $r_0$ is then implicitly defined by
\bes
  r^2 F(r)|_{r=r_0} = 1.65 \enspace . \label{e_r0}
\ees
On the other hand, to obtain a phenomenological value for $r_0$,
one needs to assume an approximate validity of potential models
for the description of the spectra  of
$c \bar c$ and $b \bar b$ mesons. 
This is illustrated in \fig{f_def_r0}.
In fact, the value $1.65$ on the r.h.s. of \eq{e_r0} has been chosen
to have $r_0=0.5\,\fm$ from the Cornell potential. This is
a distance which is well 
within the range where the observed bound states determine (approximately)
the phenomenological potential.
\begin{figure}[ht]
\vspace{-3.5cm}

\psfig{file=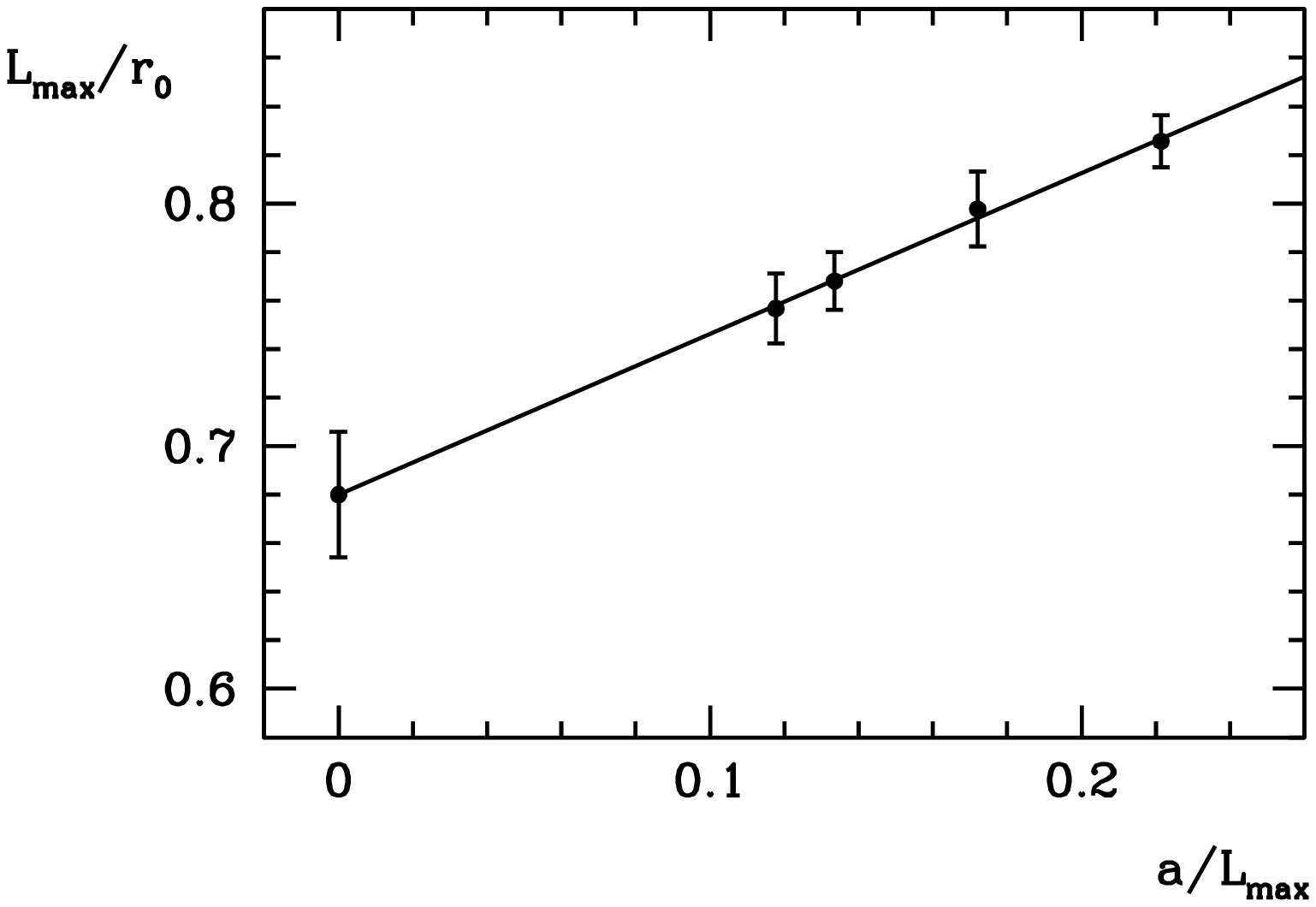,width=11cm}
\vspace{-1.5cm}
\protect\caption{Continuum extrapolation of $\Lmax/r_0$, using data
of  \protect\cite{UKQCDr0,alphaIII}.
 \protect\label{f_ml_r0}}
\end{figure}

In the following we set $r_0=0.5\,\fm$, emphasizing that this is mainly 
for the purpose of illustration and should be replaced 
by a direct experimental observable once one computes the coupling
in full QCD.

To obtain $\Lmax/r_0$ from lattice QCD,
one picks a certain value of $L/a$, tunes the bare coupling $g_0$ such
that $\gbar^2=3.48$. At the same value of $g_0$ one then computes
the force $F(r)$ on a lattice that is large enough such that finite size 
effects are negligible for the calculation of $F(r)$ and determines $r_0$. 
Repeating the calculation
for various values of $L/a$ one may extrapolate the lattice results to
zero lattice spacing (\fig{f_ml_r0}) and can quote the energies
$q$ in $\GeV$, as done in \fig{f_running}.

\subsection{Matching at finite energy \label{s_Mfe}}

Following the strategy of \fig{f_strategy},
one finally 
computes the $\Lambda$-parameter in the SF scheme. It may be converted 
to any other scheme through a 1-loop calculation. There is no perturbative
 error in 
this relation, as the $\Lambda$-parameter refers to infinite
energy, where $\alpha$ is arbitrarily small. 

Nevertheless, in order to clearly explain the problem,
we first consider 
changing schemes perturbatively
at a finite but large value of the energy.
Before writing down the perturbative relation between
$\alpha_X$ and $\alpha_Y$ where $X,Y$ label the schemes,
we note that in any scheme, there is an ambiguity 
in the energy scale $q$ used as argument for $\alpha$. 
For example
in the SF-scheme, we have set $q=1/L$, but  a choice $q=\pi/L$ 
would have been possible as well. 
This suggests immediately to allow for the freedom 
to compare the couplings after a relative energy shift. 
So we introduce a scale
factor $s$ in the perturbative relation, 
\bes
 \alpha_Y({ s}q) = \alpha_X(q) + c_1^{XY}({ s}) [\alpha_X(q)]^2 +
       c_2^{XY}({ s}) [\alpha_X(q)]^3 + \ldots \enspace .
       \label{e_match}
\ees
A natural and non-trivial question is now, 
which  scale ratio $s$ is optimal. A possible criterion is to 
choose $s$ such that the available terms in the perturbative 
series \eq{e_match} are as small as possible.  Since the number of 
available terms  in the series is usually low, we concentrate 
here on the possibility to set the first non-trivial term to zero. 
When available, the higher order one(s) may be used to test 
the success
of this procedure. 
\begin{table}[ht]
 \centering
\begin{tabular} {|lccc|}
\hline
&&&\\[-1.0ex]
Scheme $X$  &   ~$c_1^{X~\MSbar}(1)$~   
          &   ~$c_2^{X~\MSbar}(1)$~ &   ~$c_2^{X~\MSbar}({ s_0})$~  \\
&&&\\[-1.0ex]
\hline
$ q\bar q$   & $ -0.0821$  & $-2.24$ & $-2.19$ \\
SF & $1.256$ & $2.775$ & $0.27$ \\
SF~~SU(2) & 0.943  & $1.411$ & $ 0.058$ \\
TP~~SU(2) & $-0.558$  & & \\
\hline 
 \end{tabular}
 \caption{Examples for perturbative coefficients in \eq{e_match}
          for 
          $\nf=0$.
          \label{t_match}}
\end{table}

So we fix $s$ by requiring $c_1^{XY}(s)=0$ , which is 
satisfied for $s=s_0$ with
\bes
     s_0 &=&  \exp\{-c_1^{XY}(1)/(8\pi b_0)\} = \Lambda_X/\Lambda_Y ,
     \label{e_s0}
\ees
a relative shift given by the ratio of the $\Lambda$-parameters in 
the two schemes. Examples taken from
the literature
\myref{alphaII,alphaIII,twoloop1,twoloop2,twoloop3,qqbar1loop1,oneloopferm,qqbar1loop2,qqbar2loop} 
are listed in \tab{t_match}. 
In the case of matching the SF-scheme to $\MSbar$,
the use of $s_0$ does indeed reduce the 2-loop coefficient 
considerably. However for the $q\bar{q}$-scheme $s_0$ is close to
one and the 2-loop coefficient remains quite big. 
Not too surprisingly, no universal success of \eq{e_s0}
is seen.

\begin{figure}[ht]
\vspace{-7cm}

\psfig{file=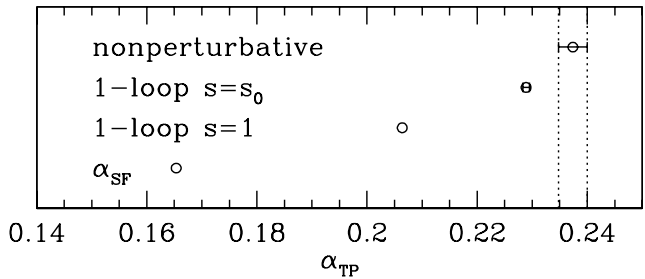,%
width=11cm}
\vspace{-0.5cm}
\caption{Non-perturbative test of perturbative matching.
        \label{f_match_sf_tp}}
\end{figure}

A non-perturbative test of the perturbative matching has been
carried out by \cite{alphaTPSF} in the SU(2) Yang-Mills
theory, where the SF-scheme was related to a different finite volume
scheme, called TP.\footnote{
For the definition of the TP-scheme
we refer the reader to the 
literature~\myref{alphaTPI,alphaTPII}.}
The matching coefficient for this case is also listed in \tab{t_match}. 
Non-perturbatively the matching was computed as follows.
\begin{itemize}
\item {For fixed $L/a$, 
       the bare coupling was tuned such that
       $\gbar^2_{\rm SF}(L)=2.0778$ (or equivalently $\alpha_{\rm SF}(q=1/L)=0.1653$)}.
\item {At the same bare coupling $\gbar^2_{\rm TP}(L)$ was computed.} 
\item {These steps were repeated for a range of $a/L$ and
       the results for $\gbar^2_{\rm TP}(L)$ were 
       extrapolated to the continuum.}
\end{itemize}
The result is shown in \fig{f_match_sf_tp}. 

We observe that a naive application of the 1-loop formula with $s=1$
falls far short of the non-perturbative number (the point with error bar),
while inserting $s=s_0$ gives a perturbative estimate
which is close to the true answer.
Indeed, the left over difference is roughly of a magnitude $\alpha^3$.

Nevertheless, without the non-perturbative result, the error inherent
in the perturbative matching is rather difficult to estimate.
For this reason it is very attractive to perform the matching at infinite 
energies, i.e. through the $\Lambda$--parameters,
where no perturbative error remains. 

\subsection{The $\Lambda$ parameter of quenched QCD}

We first note that the $\Lambda$-parameter in a given scheme
is just the integration constant in the solution of the 
renormalization group equation. This is expressed by the exact relation
\bes
 \Lambda &=&q \left(b_0\gbar^2\right)^{-b1/(2b_0^2)} \rme^{-1/(2b_0\gbar^2)}
           \exp \left\{-\int_0^{\gbar} \rmd x 
          \left[\frac{1}{ \beta(x)}+\frac{1}{b_0x^3}-\frac{b_1}{b_0^2x}
          \right]
          \right\} \enspace . \label{e_lambdapar}
\ees
We may evaluate this expression for the last few data points
in \fig{f_running} using the 3-loop approximation to the 
$\beta$-function in the SF-scheme. 
The resulting $\Lambda$-values are essentially independent 
of the starting 
point, since the data follow the perturbative running very accurately.
This excludes a sizeable contribution to the $\beta$-function beyond
3-loops and indeed, a typical estimate of a 4-loop term 
in the $\beta$-function would change the value of $\Lambda$ 
by a tiny amount. The corresponding uncertainty can be neglected compared
to the statistical errors.

After converting to the $\MSbar$-scheme one arrives at the 
result~\myref{lat97}
\bes
 \Lambda_{\MSbar}^{(0)} = 251 \pm 21 \MeV \enspace, 
 \label{e_lambdares}
\ees
where the label $^{(0)}$ reminds us that this number 
was obtained with zero quark flavors, i.e. in the Yang-Mills theory. 
Since this is not the physical theory, one must also remember that 
the overall scale of the theory was set by putting $r_0=0.5\,\fm$. 
We emphasize that the error in \eq{e_lambdares} sums up all 
errors including the extrapolations to the continuum limit 
that were done in the various intermediate steps.

\subsection{The use of bare couplings \label{s_bc}}

As mentioned before, the recursive finite size technique has 
not yet been applied to QCD with quarks. Instead,
$\alpha_{\MSbar}$ has been estimated through lattice gauge
theories by using a short cut, namely the relation between
the bare coupling of the lattice theory and the $\MSbar$-coupling
at a physical momentum scale which is of the order of
the inverse lattice spacing that corresponds to the
bare coupling~\myref{fermilab}. 
Without going too much into details, 
we want to discuss this approach, its merits and its shortcomings,
here. The emphasis is on the principle and not on the applications,
which can be found in \cite{alpha_dyn}. So, although the
main point is to be able to include quarks, we set $\nf=0$
in the discussion; more is known in this case!

The method simply requires that one computes one dimensionful 
experimental observable in lattice QCD at a certain value of the
bare coupling $g_0$. A popular choice for this is a mass splitting
in the $\Upsilon$-system \myref{Davies}. Using as input the experimental 
mass splitting one determines the lattice spacing in physical units.

Next one may attempt to use the perturbative relation, 
\bes
 \alphaMSbar(s_0 a^{-1}) = \alpha_0 + { 4.45} \alpha_0^3 + \rmO(\alpha_0^4)
  + \Oa \, , \quad \alpha_0= g_0^2/(4\pi) \label{e_alpha0}
\ees
to get an estimate for $\alphaMSbar$.
Here we have already inserted a scale shift $s_0$ (cf. \sect{s_Mfe}). 
Without this 
scale shift, the 1-loop and 2-loop coefficients in the above equation would 
be very large. In turn this means that the shift, 
\bes
s_0=28.8 \enspace ,
\ees
is enormous. Furthermore, the series \eq{e_alpha0}
does not look very healthy
even after employing $s_0$. 
Such a behavior of power expansions in $\alpha_0$ 
has also been observed for other 
quantities~\myref{Lepenzie}. One concludes that $\alpha_0$ is a bad 
expansion parameter
for perturbative estimates.

The origin of this problem appears to be a 
large renormalization between the bare coupling and 
general observables defined at the scale of the lattice cutoff
$1/a$. Assuming this large renormalization to be roughly universal, 
one
can cure the problem by inserting the non-perturbative (MC) 
values of a 
short distance observable~\myref{Parisi,Lepenzie}, the obvious candidate being 
\bes
 P=\frac1N \langle \tr U(p) \rangle  \enspace . 
\ees
In detail, due to the
perturbative expansion,
\bes
 - \frac{1}{\cf \pi} \ln(P)= \alpha_0 + 3.373  \alpha_0^2 + 17.70 \alpha_0^3+
 \ldots \enspace ,
\ees
we may define an {\it improved bare coupling},
\bes
 \alpha_{\Box} \equiv  - \frac{1}{\cf \pi} \ln(P) \enspace , 
  \label{e_alphabox}
\ees
which appears to have a regular perturbative relation to
$\alphaMSbar$,
\bes
 \alphaMSbar(s_0 a^{-1}) = \alpha_{\Box} + {0.614} \alpha_{\Box}^3 
 + \rmO(\alpha_{\Box}^4) + \Oa \enspace .
 \label{e_alpha_MS_box}
\ees
Of course, the point of the exercise is to insert the average, $P$, obtained 
in the MC calculation into \eq{e_alphabox}. Afterwards one only needs
to use the (seemingly) well behaved expansion \eq{e_alpha_MS_box}.
One can construct many other improved bare couplings but the
assumption is that the aforementioned large renormalization of
the bare coupling is roughly universal and the details do not matter 
too much.

On the one hand, the advantages of \eq{e_alpha_MS_box} are obvious: 
i) one only needs
         the calculation of a hadronic scale and ii)
         the 2-loop relation to $\alphaMSbar$ is known (for $n_f=0$).
On the other hand, how was the problem of scale dependent renormalization
(\sect{s_Rls}) solved? 
It was not! To remind us, the general problem
is to reach large energy scales,
where perturbation theory may be used in a controlled way. 
In the present context this would require to compute with a
series of lattice spacings for which  $\alphaMSbar(s_0 a^{-1})$ 
is both small and changes
appreciably. The required lattice sizes would then be too large to perform 
the calculation. Therefore one must {\it assume} that the error
terms in \eq{e_alpha_MS_box} are small. A particular worry is that 
one may  not take
the continuum limit -- due to the very nature of \eq{e_alpha_MS_box}, 
which says that $\alpha$ runs with the lattice spacing. 
This means that
it is impossible to disentangle the $\Oa$ and
the $\rmO(\alpha_{\Box}^4)$ errors.

\begin{figure}
\vspace{-2.1cm}

\centerline{
\psfig{file=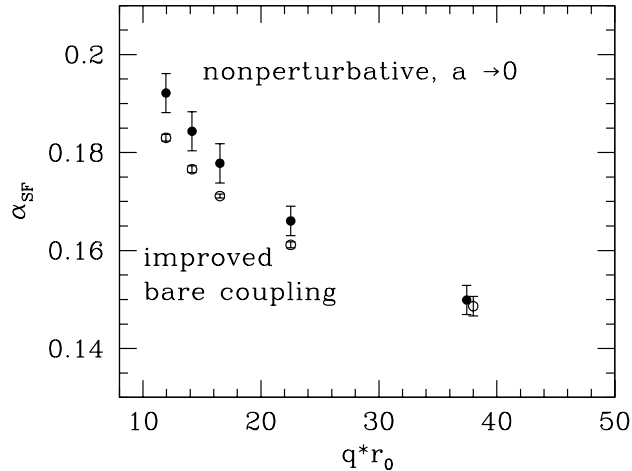,%
width=9cm}}
\vspace{-1.0cm}
\caption{Test of an improved bare coupling in the SU(2) 
Yang-Mills theory.
\label{f_alpha_plaq_su2}}
\end{figure}

We briefly demonstrate now that this last 
worry is justified in
practice. For this purpose we consider the SU(2) Yang-Mills theory,
where  $\alpha_{\rm SF}$ was computed non-perturbatively 
and in the continuum limit, as a function of the energy scale $q$
in units of $r_0$ \myref{alphaTPSF}. The results of this computation 
are shown as points with error bars in \fig{f_alpha_plaq_su2}. 
We may now compare them to the estimate in terms of the
improved bare coupling,
\bes
        \alpha_{\rm SF}(q)&=& \alpha_{\Box} + {0.231} \alpha_{\Box}^3
        \, , \quad q={s_0} a^{-1}  \, , \quad {s_0}=1.871 \enspace ,
\ees
where the only inputs needed are $P$ as well as the
value of $r_0/a$ since $q r_0 = {s_0} r_0/a$. These estimates are given
as circles in the figure.

In general, and in particular for large values of $qr_0$, 
the agreement is rather good. However, for the lower values of
$qr_0$, significant differences are present, which are far underestimated
by a perturbative error term $\alpha^4$. 

What does this teach us about the method as applied in full QCD? To this end, 
we note that the lattice spacings that are used in the
applications of improved bare couplings in full QCD calculations,
correspond to $q*r_0 < 15$. This is the range where we saw significant 
deviations in our test.
In light of this
it appears 
to us that the errors that are usually quoted for $\alphaMSbar$ using this 
method are underestimated. It is encouraging, though, that the values 
which are obtained in this way
compare well with those extracted from experiments using 
other methods~\myref{alpha_dyn}.

\section{Renormalization group invariant quark mass}

The computation of running quark masses and the 
renormalization group invariant (RGI) quark mass \myref{lat97} proceeds in
complete analog to the computation of $\alpha(q)$.
Since we are using a mass-independent renormalization scheme (cf. \sect{s_Rm}),
the renormalization
(and thus the scale dependence) is independent of the flavor of
the quark.  When we consider ``the'' running mass below, any one flavor 
can be envisaged; the scale dependence is the same for all of them.

The renormalization group equation for the coupling \eq{e_RG}
is now accompanied
by one describing the scale dependence of the mass,
\bes
  q {\partial \mbar \over \partial q} &=& \tau(\bar g) \enspace ,
     \label{e_RG_m}
\ees
where $\tau$ has an asymptotic expansion
\bes     
 \tau(\bar g) & \buildrel {\bar g}\rightarrow0\over\sim &
 -{\bar g}^2 \left\{ d_0 + {\bar g}^{2}  d_1 + \ldots \right\}
                      \, , \qquad
 d_0={8}/{(4\pi)^2} 
 \enspace ,  \label{e_RGpert_m}
\ees
with higher order coefficients $d_i, \, i>0 $ which depend on the scheme.

Similarly to the $\Lambda$-parameter, we may define a 
renormalization group invariant quark mass, $M$, by
the asymptotic behavior of $\mbar$,
\bes
 M &=& \lim_{q \to \infty} \mbar (2 b_0\gbar^2)^{-d_0/2b_0^2}\enspace .
\ees
It is easy to show that $M$ does not depend on the renormalization scheme. 
It can be computed in the SF-scheme and used
afterwards to obtain the running mass in any other scheme by inserting
the proper $\beta$- and $\tau$-functions in the renormalization
group equations.

\begin{figure}[ht]
\vspace{ -3.0cm}

\centerline{
\psfig{file=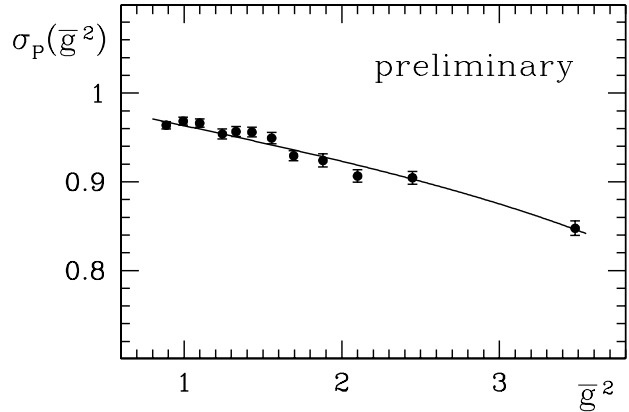,%
width=11cm}}
\vspace{-2.3cm}
\caption{The step scaling function for the quark mass.
\label{f_sigma_p2_fit}}
\end{figure}
To compute the scale evolution of the mass non-perturbatively, 
we introduce a new step scaling function,
\bes
 \sigmap = \zp(2L) / \zp(L) \enspace .
\ees
The definition of the corresponding lattice
step scaling function and the extrapolation to the continuum
is completely analogous to the case of $\sigma$. The only
additional point to note is that one needs to keep 
the quark mass zero throughout the calculation. This is achieved
by tuning the bare mass in the lattice action
such that the PCAC mass \eq{e_PCAC} vanishes. At least in the quenched 
approximation, which has been used so far, this turns out to be
rather easy  \myref{paper3}.

First results for $\sigmap$ (extrapolated to the continuum) have
been obtained recently~\myref{lat97}. They are displayed in 
\fig{f_sigma_p2_fit}.
\begin{figure}[ht]
\vspace{ -2.0cm}

\centerline{
\psfig{file=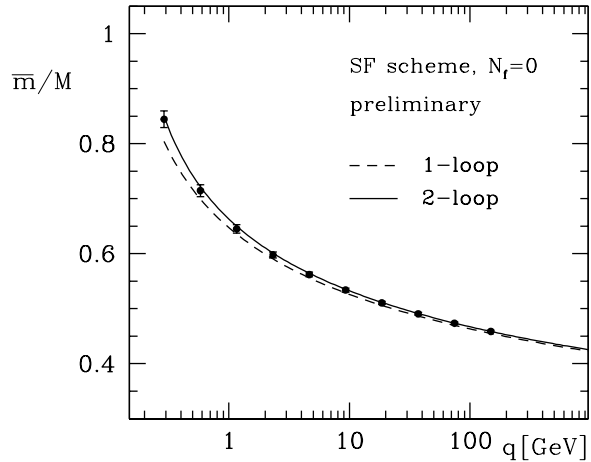,%
width=10cm}}
\vspace{-2.0cm}
\caption{The running quark mass as a function of $q\equiv1/L$.
\label{f_mbar}}
\end{figure}
 
Applying $\sigmap$ and $\sigma$ recursively one then obtains
the series, 
\bes
 \mbar(2^{-k} \Lmax) / \mbar(2\Lmax)\, , \,\,k=0,1,\ldots \enspace ,
\ees  
up to a largest value of $k$, which corresponds to the smallest
$\gbar$ that was considered in \fig{f_sigma_p2_fit}.  
From there on,
the perturbative
2-loop approximation to the $\tau$-function and
3-loop approximation to the $\beta$-function
(in the SF-scheme) 
may be used to integrate the renormalization group equations to infinite
energy, or equivalently  to $\gbar=0$. The result is the
renormalization group invariant mass,
\bes
  M = \mbar\,(2 b_0\gbar^2)^{-d_0/2b_0^2} 
   \exp \left\{- \int_0^{\gbar} \rmd g [\frac{\tau(g)}{\beta(g)}
     - \frac{d_0}{b_0 g} ] \right\}  \enspace .
\ees
In this way, one is finally able to express the
running mass $\mbar$ in units of the renormalization group invariant mass, 
$M$, as shown in \fig{f_mbar}. $M$ has the same value in all
renormalization schemes, in contrast to the
running mass $\mbar$.  

The perturbative evolution is again very accurate down to
low energy scales. Of course,
this result may not be 
generalized 
to running masses in other schemes. Rather the running  has to be 
investigated in each scheme separately.

The point at lowest energy in \fig{f_mbar} corresponds to
\bes
 M/\mbar = 1.18(2) \quad \mbox{at} \quad L=2\Lmax \enspace .
\ees
Remembering the very definition of the renormalized mass \eq{e_mbar},
one can use this result to relate the renormalization group invariant mass
mass and
the bare current quark mass $m$ on the lattice
through
\bes
  M= m \times 1.18(2) \times \za(g_0) / \zp(g_0,2\Lmax/a) \enspace . 
  \label{e_M_final}
\ees
In this last step, one should insert the bare current quark mass, e.g. of
the strange quark, and extrapolate the result to the continuum limit.
This analysis has not been finished yet but results including this last step 
are to be expected, soon. 
To date, the one-loop approximation for the renormalization
of the quark mass (i.e. an approach similar to what was discussed
for the coupling in~\sect{s_bc}) has been used to obtain numbers for
the strange quark mass in the $\MSbar$-scheme. 
The status of these determinations was recently reviewed 
by \cite{review_mstrange}.

\section{Chiral symmetry, normalization of currents and $\Oa$-improvement
         \label{s_curr}}

In this section we discuss two renormalization problems that are of 
quite different nature. The first one is the renormalization of
irrelevant operators, that are of interest in the systematic
$\Oa$ improvement of Wilson's lattice QCD as mentioned in \sect{s_Io}.
The second one is the finite normalization of isovector currents (cf.
\sect{s_finiter}).
They are discussed together, here,  because -- at least to a
large extent -- they can be treated with a proper application 
of chiral Ward identities. The possibility to use
chiral Ward identities to normalize the currents has first been discussed
by \cite{Boch,MaMa}. Earlier numerical
applications can be found in 
\cite{MartinelliEtAlI,PacielloEtAl,HentyEtAl} and a complete calculation
is described below~\myref{paper4}. 
We also sketch the application
of chiral Ward identities in the
computation of the $\Oa$-improved
action and currents~\myref{paper1,paper2,paper4}.

Before going into the details, we would like to convey the rough idea
of the application of chiral Ward identities. 
For simplicity we again assume an isospin doublet of mass-degenerate 
quarks. 
Imagine that we have
a regularization of QCD which preserves the full
SU($2)_{\rm V} \times $SU($2)_{\rm A}$
flavor symmetry as it is
present in the continuum Lagrangian of mass-less QCD.
In this theory we can derive chiral Ward identities,
e.g. in the Euclidean formulation of the theory. These 
then provide exact relations between different correlation functions.
Immediate consequences of these relations are that the 
currents \eq{e_currents} do not get renormalized ($\za=\zv=1$)
and the quark mass does not have an additive renormalization.

Lattice QCD does, however, not have the full
SU($2)_{\rm V} \times $SU($2)_{\rm A}$ flavor symmetry for finite
values of the lattice spacing and in fact no regularization is known
that does. Therefore, the Ward identities 
are not satisfied exactly. We do, however, expect that the 
renormalized correlation functions obey the same Ward identities
as before -- up to $\Oa$ corrections that vanish in the continuum
limit. Therefore we may impose those Ward identities for
the renormalized currents, to fix their
normalizations. 

Furthermore, following Symanzik, it suffices to a add
a few local irrelevant terms to the action and to the currents
in order to obtain an improved lattice theory, where the continuum
limit is approached with corrections of order $a^2$.
The coefficients of these terms can be determined 
by imposing improvement conditions. For example one may 
require certain chiral Ward identities to be valid at finite lattice
spacing $a$. 

\subsection{Chiral Ward identities}

For the moment we do not pay attention to a regularization of
the theory and derive the Ward identities in a formal way. As mentioned
above these identities would be exact in a regularization that preserves
chiral symmetry. To derive the Ward identities, one starts from the
path integral representation of a correlation function 
and performs the change of integration variables
\bes
  \psi(x) &\to& \rme^{i \frac{\tau^a}{2}
            [\eps^a_{\rm A}(x)\gamma_5 + \eps^a_{\rm V}(x)]} \psi(x) 
    \nonumber \\
         &=& \psi(x) + i  \eps^a_{\rm A}(x)   \da^a\psi(x)
         + i \eps^a_{\rm V}(x) \dv^a\psi(x)\enspace ,
    \nonumber \\ 
  \psibar(x) &\to& \psibar(x) \rme^{i \frac{\tau^a}{2}
            [\eps^a_{\rm A}(x)\gamma_5 - \eps^a_{\rm V}(x)]}  
    \nonumber \\
         &=& \psibar(x) + i  \eps^a_{\rm A}(x)   \da^a\psibar(x)
         + i \eps^a_{\rm V}(x) \dv^a\psibar(x) \enspace ,
\ees
where we have taken $\eps^a_{\rm A}(x), \eps^a_{\rm V}(x)$ infinitesimal
and introduced the variations
\bes
  \dv^a\psi(x)&=&\frac{1}{2}\tau^a\psi(x),
  \qquad\quad\quad
  \dv^a\psibar(x)=-\psibar(x)\frac{1}{2}\tau^a \enspace , 
  \nonumber \\
  \da^a\psi(x)&=&\frac{1}{2}\tau^a\dirac{5}\psi(x),
  \qquad\phantom{\dirac{5}}
  \da^a\psibar(x)=\psibar(x)\dirac{5}\frac{1}{2}\tau^a \enspace .
  \label{e_var_quarks}
\ees
The Ward identities then follow from the invariance of the
path integral representation of correlation functions with respect to such
changes of integration variables. They obtain contributions from
the variation of the action and the variations of the fields
in the correlation functions. In \sect{s_Nic} we will need 
the variations of the currents,
\bes
  \dv^aV^b_{\mu}(x)&=&-i\epsilon^{abc}V^c_{\mu}(x),
  \qquad\,
  \da^aV^b_{\mu}(x)=-i\epsilon^{abc}A^c_{\mu}(x),
  \nonumber \\
  \dv^aA^b_{\mu}(x)&=&-i\epsilon^{abc}A^c_{\mu}(x),
  \qquad
  \da^aA^b_{\mu}(x)=-i\epsilon^{abc}V^c_{\mu}(x) \enspace .
  \label{e_var_currents}
\ees
They form a closed algebra under these variations.

Since this is convenient for our applications, we write the
Ward identities in an integrated
form. 
Let $R$ be a space-time region with smooth boundary $\partial R$.
Suppose ${\cal O}_{\rm int}$ and ${\cal O}_{\rm ext}$ 
are polynomials in the basic fields localized
in the interior and exterior of $R$ respectively.
The general vector current Ward identity then reads
\bes
  \int_{\partial R}\rmd\sigma_{\mu}(x)\,  
  \left\langle 
  V^a_{\mu}(x) {\cal O}_{\rm int} {\cal O}_{\rm ext} 
  \right\rangle
  =-
  \left\langle 
  \left(\dv^a{\cal O}_{\rm int}\right) {\cal O}_{\rm ext} 
  \right\rangle,
  \label{e_vectorWI}
\ees
while for the axial current one obtains
\bes
  \int_{\partial R}\rmd\sigma_{\mu}(x)\,  
  \left\langle 
  A^a_{\mu}(x) {\cal O}_{\rm int} {\cal O}_{\rm ext} 
  \right\rangle
  &=&-
  \left\langle 
  \left(\da^a{\cal O}_{\rm int}\right) {\cal O}_{\rm ext} 
  \right\rangle  
  \label{e_axialWI} \\
  &&+
  2m\int_R\rmd^4x\,
  \left\langle 
  P^a(x) {\cal O}_{\rm int} {\cal O}_{\rm ext} 
  \right\rangle \enspace .
  \nonumber
\ees
The integration measure $\rmd\sigma_{\mu}(x)$ points 
along the outward normal to the surface $\partial R$
and the pseudo-scalar density $P^a(x)$ is defined by
\bes
  P^a(x)=\psibar(x)\dirac{5}\frac{1}{2}\tau^a\psi(x) \enspace .
\ees

We may also write down  the precise meaning of the PCAC-relation \eq{e_PCAC}. 
It is \eq{e_axialWI} in a differential form, 
\bes
 \left\langle \left[ \partial_{\mu}A_{\mu}^a(x) -2m
  P^a(x)\right]   {\cal O}_{\rm ext} 
  \right\rangle = 0 \enspace , 
  \label{e_PCACnew}
\ees
where now ${\cal O}_{\rm ext}$ may have support everywhere but at the point
$x$. 

Going through the same derivation in the lattice regularization,
one finds equations of essentially  the same form as the 
ones given above, but with additional terms~\myref{Boch}.
At the classical level these terms are of order $a$. More precisely, 
in \eq{e_PCACnew} the important additional term
originates from the variation of the Wilson term,
$a\,\psibar \nabstar{\mu} \nab{\mu} \psi$, and is a local field of dimension 5.
Such $\Oa$-corrections are present in any observable computed on the lattice
and are no reason for concern. However, as is well known in field theory,
such operators mix with the ones of lower and equal dimensions
when one goes 
beyond the classical approximation.
In the present case, the dimension
five operator mixes amongst others also with 
$\partial_{\mu}A_{\mu}^a(x)$ and $P^a(x)$.
This means that part of the classical $\Oa$-terms turn into
$\rmO(g_0^2)$ in the quantum theory. The 
essential observation is now that this mixing 
can simply be written in the form of a
renormalization of the terms that are already present
in the Ward identities, since all dimension three and four operators 
with the right quantum number are already there.

We conclude 
that the identities, which we derived above in a formal manner, 
are valid in the lattice regularization after 
\begin{itemize}
 \item{replacing the bare fields $A,V,P$ and quark mass $m_0$
 by renormalized ones, where one must allow 
 for the most general
 renormalizations,
 \bes
  (\ar)_{\mu}^a &=& \za A_{\mu}^a \enspace ,\quad
 (\vr)_{\mu}^a = \zv V_{\mu}^a \enspace , \nonumber \\
 (\pr)^a &=& \zp P^a \enspace , \quad
 \mr = \zm \mq\, , \quad \mq=m_0-m_c  \enspace , \nonumber
\ees}
 \item{allowing for the usual $\Oa$ lattice artifacts.}
\end{itemize}
Note that the additive quark mass renormalization $m_c$ 
diverges like $\rmO(g_0^2/a)$  for dimensional
reasons.  

As a result of this discussion, the formal
Ward identities may
be used to determine the normalizations of the currents.
We discuss this in more detail in \sect{s_Nic} and first explain
the general idea how one can use the Ward identities to determine
improvement coefficients.

\subsection{$\Oa$-improvement}

We  refer the reader to \cite{paper1} for a thorough discussion
of $\Oa$-improvement and to \cite{Sommer97} for a review. 
Here, we only sketch how chiral Ward identities 
may be used to 
determine  improvement coefficients non-perturbatively. 

The form of the improved action and the improved
composite fields is determined by the symmetries
of the lattice action and in addition the equations of motion 
may be used to reduce the set of operators that have to be
considered \myref{OnShell}. For $\Oa$-improvement, the improved action
contains only one additional term, which is conveniently chosen 
as \myref{SW}
\bes
 \delta S = a^5 \sum_x \csw \psibar(x)\frac i4 \sigma_{\mu\nu}
  \widehat{F}_{\mu\nu}(x) \psi(x),
\ees
with $\widehat{F}_{\mu\nu}$ a lattice approximation to the gluon field 
strength tensor ${F}_{\mu\nu}$ and 
one improvement coefficient
$\csw$.
The improved and renormalized currents 
may be written in the general form
\bes 
(\vr)_\mu^a &=&
 \zv(1+\bv a\mq)\bigl\{V_\mu^a+a\cv
               \frac12({\partial}_\nu +{\partial}^*_\nu)
       T_{\mu \nu}^a\bigr\},\nonumber\\
       && T_{\mu \nu}^a(x) = i\psibar\sigma_{\mu\nu}\frac{1}{2}\tau^a\psi(x)
       \nonumber\\
(\ar)_\mu^a &=&
 \za(1+\ba a\mq)\bigl\{A_\mu^a+a\ca
               \frac12({\partial}_\mu +{\partial}^*_\mu)
       P^a\bigr\},\nonumber\\
 (\pr)^a &=&
 \zp(1+\bp a\mq)P^a\, . \label{e_renfields}
\ees
(${\partial}_\mu$ and  ${\partial}^*_\mu$ are the forward and backward
lattice derivatives, respectively.)

Improvement coefficients like $\csw$ and $\ca$ are functions of the 
bare coupling, $g_0$,
and need to be fixed by suitable
improvement conditions. One considers pure lattice artifacts,
i.e. combinations
of observables that are known to vanish in the continuum limit
of the theory. Improvement conditions require
these lattice artifacts to vanish,
thus defining the values of the improvement 
coefficients as a function of the lattice spacing (or equivalently as a 
function of $g_0$).

In perturbation theory, lattice artifacts 
can be obtained from any (renormalized) quantity by subtracting its
value in the continuum limit. The improvement coefficients are 
unique.

Beyond perturbation theory, one wants to determine the improvement coefficients
 by
MC calculations and it requires significant effort
to take the continuum limit.
It is therefore advantageous to use lattice artifacts that derive from
a symmetry of the continuum field theory that is not respected by the
lattice regularization. One may require rotational invariance
of the static potential $V({\bf r})$, e.g.
$$
 V({\bf r}=(2,2,1)r/3) - V({\bf r}=(r,0,0)) =0 \, ,
$$
or 
Lorenz invariance,
$$
 [E({\bf p})]^2 - [E({\bf 0})]^2  - {\bf p}^2 =0 \, ,
$$
for the momentum dependence of a one-particle energy $E$.

For $\Oa$-improvement of QCD it is advantageous
to require instead that particular
chiral Ward identities are valid {\it exactly}.\footnote{
 As a consequence of the freedom to choose improvement conditions, 
the resulting values of 
improvement coefficients such as $\csw,\,\ca$
depend on the exact choices made. The corresponding variation of
$\csw,\,\ca$ is of order $a$. 
There is nothing wrong with this unavoidable fact, since
an $\Oa$ variation in the improvement
coefficients only changes the effects of
order $a^2$ in physical observables computed after improvement.}

In somewhat more detail, the determination of $\csw$ and $\ca$
is done as follows. 
We define a bare current quark mass, $m$, viz.
\bes
 m \equiv {  \left\langle \left[ \partial_{\mu}(\aimpr)_{\mu}^a(x) \right]
              {\cal O}_{\rm ext} \right\rangle 
           \over
             2 \left\langle  P^a(x)   {\cal O}_{\rm ext} 
                 \right\rangle }
                 \, , \quad 
   (\aimpr)_\mu^a=A_\mu^a+a\ca \frac12({\partial}_\mu +{\partial}^*_\mu)
                  P^a \enspace . 
  \label{e_PCACnn}
\ees
When all improvement coefficients have their proper values, 
the renormalized quark mass, defined by
the renormalized PCAC-relation, is related
to $m$ by
\bes
  \mr = {{\za (1+\ba a \mq)}\over{\zp (1+\bp a \mq)}} m + \rmO(a^2)\enspace .
  \label{e_mrn}
\ees
We now choose 3 different versions of \eq{e_PCACnn} by different
choices for ${\cal O}_{\rm ext}$ and/or boundary conditions and
obtain 3 different values of $m$, denoted by $m_1,m_2,m_3$.
Since the prefactor in front of $m$ in \eq{e_mrn} is just a numerical 
factor, we may conclude that all $m_i$ have to be equal
in the improved theory up to errors of order $a^2$.
$\csw$ and $\ca$ may therefore be computed by requiring
\bes
  m_1 = m_2 = m_3 \label{e_imprcond}\enspace .
\ees
This simple idea has been used to compute 
$\csw$ and $\ca$ as a function of $g_0$ in the quenched 
approximation~\myref{paper3}. 
In the theory with two flavors of dynamical
quarks, $\csw$ has been computed in this way~\myref{csw_dyn}.
The improvement coefficient for 
the vector current, $\cv$,
 may be computed through a different chiral Ward identity~\myref{cv}.

\subsection{Normalization of isovector currents \label{s_Nic}}

Although the numerical results, which we will show below,
have been obtained after $\Oa$ improvement,
the normalization of the currents as it is
described, here,  
is applicable in general. Without improvement one just 
has to remember that the error terms are of order $a$, instead of $a^2$.
For the following, 
we set the quark mass (as calculated from the PCAC-relation) to zero.

\subsubsection{Normalization condition for the vector current.}

Since the isospin symmetry of the continuum theory is preserved on 
the lattice exactly,
there exists also an exactly conserved vector current. This means
that certain specific Ward-identities for this current are
satisfied exactly and fix it's normalization automatically.
It is, however, convenient to use the improved vector current introduced above,
which is only conserved up to cutoff effects of order $a^2$.
Its normalization is hence not naturally given and we 
must impose a normalization condition to fix $\zv$.
Our aim in the following is to derive such a condition 
by studying the action of the renormalized isospin charge 
on states with definite isospin quantum numbers.

The matrix elements that we shall consider are constructed in the 
SF using (the lattice version of) the boundary field products
introduced in \eq{e_boundops} 
to create initial and final states that transform according  
to the vector representation of the exact isospin symmetry.
The correlation function 
\bes
  \fvr(x_0)={a^3\over6L^6}\sum_{\bf x}i\epsilon^{abc}
  \langle\opprime{}^a(\vr)_0^b(x)\op{}^c\rangle
\ees
can then be interpreted as a matrix element of the renormalized isospin
charge between such states. The properly normalized
charge generates an infinitesimal
isospin rotation and 
after some algebra one finds that the correlation function must be equal to
\bes
  f_1=-{1\over3L^6}
  \langle\opprime{}^a\op{}^a\rangle \enspace 
  \label{e_norm_vect}
\ees
up to corrections of order $a^2$.
The O($a$) counter-term appearing in the definition \eq{e_renfields} of the 
improved
vector current does not contribute to the correlation function
$\fvr(x_0)$. So if we introduce the analogous correlation function
for the bare current,
\bes
  \fv(x_0)={a^3\over6L^6}\sum_{\bf x}i\epsilon^{abc}
  \langle\opprime{}^aV_0^b(x)\op{}^c\rangle,
\ees
it follows from \eq{e_norm_vect} that 
\bes
  \zv \fv(x_0)=f_1+\rmO(a^2).
  \label{e_norm_vect2}
\ees
By evaluating the correlation functions $f_1$ and $\fv(x_0)$
through numerical simulation
one is thus able to compute the normalization factor
$\zv$.

\subsubsection{Normalization condition for the axial current.}

To derive a normalization condition for $\za$,
we consider \eq{e_axialWI} (for $m=0$)
and choose ${\cal O}_{\rm int}$ to be the axial current at some 
point $y$ in
the interior of $R$.  The resulting identity,
\bes
  \int_{\partial R}\rmd\sigma_{\mu}(x)\, 
  \left\langle 
  A^a_{\mu}(x) A^b_{\nu}(y) {\cal O}_{\rm ext} 
  \right\rangle
  =
  i\epsilon^{abc}
  \left\langle 
  V^c_{\nu}(y){\cal O}_{\rm ext} 
  \right\rangle \enspace , \label{e_axWI}
\ees
is valid for any type of boundary conditions and space-time geometry,
but we now assume Schr\"odinger functional
boundary conditions as before. A convenient choice of the region
$R$ is the space-time volume between the hyper-planes at
$x_0=y_0\pm t$. On the lattice we 
then obtain the relation\footnote{
In the context of $\Oa$-improvement
it has been important here that the fields 
in the correlation functions
are localized at non-zero distances from each other.
Since the theory is only on-shell improved, one would otherwise
not be able to say that the error term is of order $a^2$
(cf.~sect.~2 of \cite{paper1}).}
\bes
  a^3\sum_{\bf x}
  \epsilon^{abc}
  \left\langle 
  [(\ar)^a_{0}(y_0+t,{\bf x})-(\ar)^a_{0}(y_0-t,{\bf x})]
  (\ar)^b_{0}(y) {\cal O}_{\rm ext} 
  \right\rangle &&
\nonumber \\
  = 2i
  \left\langle 
  (\vr)^c_{0}(y){\cal O}_{\rm ext} 
  \right\rangle
  +\rmO(a^2) \enspace . &&\label{e_nor_ax1}
\ees

After summing over the spatial components of $y$,
and using the fact that the axial
charge is conserved at zero quark mass (up to corrections of order $a^2$),
\eq{e_nor_ax1} becomes
\bes
a^6\sum_{\bf x,y}
  \epsilon^{abc}
  \left\langle 
  (\ar)^a_{0}(x)
  (\ar)^b_{0}(y) {\cal O}_{\rm ext} 
  \right\rangle
  =
   a^3\sum_{\bf y}i
  \left\langle 
  (\vr)^c_{0}(y){\cal O}_{\rm ext} 
  \right\rangle
  +\rmO(a^2),
  \label{e_nor_ax2}
\ees
where $x_0=y_0+t$.
We now choose the field product ${\cal O}_{\rm ext}$ so that 
the function $\fvr(y_0)$ introduced previously appears on the 
right-hand side of \eq{e_nor_ax2}.
The normalization condition for the vector current \eq{e_norm_vect2}
then allows us
to replace the correlation function $\fvr(y_0)$ by $f_1$. In this way 
a condition for $\za$ is obtained~\myref{paper4}.

\subsubsection{Lattice artifacts and results.} 

\begin{figure}[ht]
\vspace{-0.9cm}

\centerline{
\psfig{file=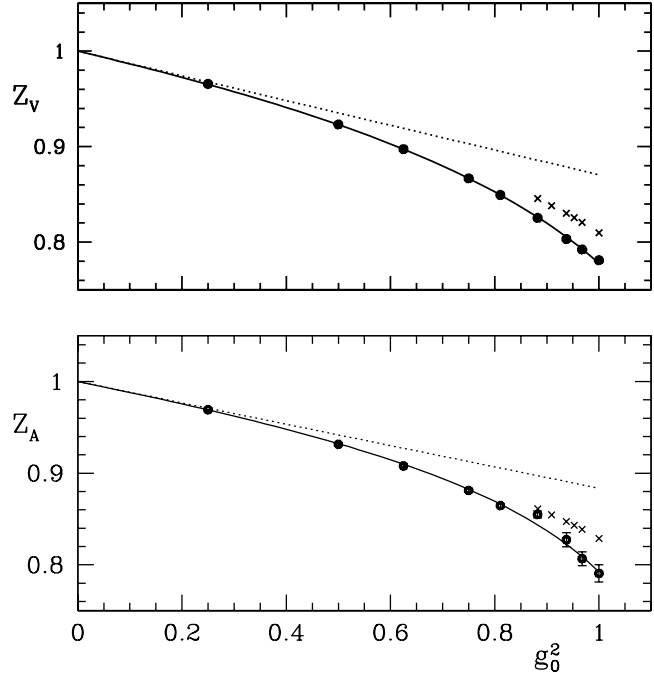,%
width=10cm}}
\vspace{0.0cm}
\protect\caption{Current normalization constants as a function
of the bare coupling~\protect\myref{paper4}.
The dotted line is 1-loop perturbation theory~\protect\myref{Za4} and the crosses
correspond to a version of 1-loop tadpole improved perturbation 
theory~\protect\myref{Lepenzie}.
The full line is a fit to the non-perturbative results.
\label{f_zvza}}
\end{figure}
It is now straightforward to compute $\zv,\za$ 
by MC evaluation of the correlation functions that enter in
\eq{e_norm_vect2},\eq{e_nor_ax2}. 
Before showing the results, we emphasize 
one point that needs to be considered carefully.
The normalization conditions fix 
$\zv$ and $\za$ only up to cutoff effects of order $a^2$.
Depending on the choice of the 
lattice size, the boundary values of the gauge field and the other
kinematical parameters that one has, slightly different results 
for $\zv$ and $\za$ are hence obtained. 
One may try to assign a 
systematic error to the normalization constants
by studying these variations in detail, but since there is no
general rule as to which choices of the kinematical parameters are 
considered to be reasonable, such error estimates are bound to be rather
subjective.

It is therefore better to deal
with this problem by {\it defining}\/ the normalization constants 
through a particular normalization condition.
The physical matrix elements of the renormalized currents 
that one is interested in must then be calculated 
for a range of lattice spacings so as 
to be able to extrapolate the data to the continuum limit.
The results obtained in this way are guaranteed to be independent
of the chosen normalization condition, because any differences 
in the normalization constants of order $a^2$ 
extrapolate to zero together with the 
cutoff effects associated with the matrix elements themselves.

Note that a ``particular normalization condition''
means that apart from choosing the boundary values and geometry
of the SF, one has to keep the size of the SF-geometry fixed in
physical units, for example
\bes
  L/r_0 = {\rm const.}
\ees 
As shown in \fig{f_zvza},
the current normalizations can be obtained with good precision
(in the quenched approximation) after taking 
all of these points into account 
\myref{paper4}. 

Coming back to our motivation \sect{s_finiter}, the results in
\fig{f_zvza} now allow for the calculation of matrix elements
of the weak currents involving light quarks
without any perturbative uncertainties.

\section{Summary, Conclusions}

We have shown how QCD needs to be renormalized non-perturbatively
in order to obtain unambiguous predictions that can be compared with 
experiments. Once a non-perturbative definition and
calculational technique is available, it is {\it in principle}
quite simple to
perform renormalization non-perturbatively. In practice, the problem has 
to be treated with care.

The only presently available definition is lattice QCD with MC simulations
as the  calculational tool to get predictions. In this case,
straightforward solutions to the renormalization problem
face a serious difficulty: the theory must be
treated at various different energy scales simultaneously,
which is an extremely hard (impossible?) task for a MC simulation.
To circumvent this difficulty, L\"uscher, Weisz and Wolff have introduced
the recursive finite size technique, where one connects low and high energies
recursively in small steps. We have shown, how this idea can be put into
practice in QCD using the Schr\"odinger functional as a second technical 
tool. In the theory without dynamical quarks, these methods
have been shown to allow for the computation of short distance parameters
like $\Lambda_{\MSbar}$ and the renormalization group invariant quark mass
with completely controlled errors! From the practical point of
view, the non-perturbative renormalization of other quantities, such as
the $\Delta s =2$ operator, still have to be investigated, but no
new difficulties are expected to appear. It is therefore plausible
that the renormalization problem can be solved for many 
specific cases -- and with good accuracy. It should not remain unmentioned,
however, that in practice each renormalization problem has to be considered
separately. Certain problems may turn out to be significantly more
difficult to solve than the ones discussed in the lectures.

We also sketched, how Symanzik improvement can be implemented in a 
non-perturbative way, reducing the leading lattice artifacts from
linear in $a$ to quadratic in $a$ ($\Oa$-improvement). 
For light quarks, such a project has already
been done carried out. As a result, significant progress 
in lattice QCD is expected
from the use of $\Oa$-improved QCD. \\[1ex]
{\bf Acknowledgment.} 
I am grateful to the organizers of this school for composing
an interesting programme and directing the school
with excellence.
I would like to thank my friends in the 
ALPHA-collaboration for all they taught me and for the most enjoyable
and fruitful collaboration; special thanks go to
J. Heitger and F. Jegerlehner for critical comments on the manuscript. 
I further thank DESY for allocating computer 
time for the ALPHA-project, which was essential for a number
of the numerical investigations that were discussed in these lectures.

%
%

\end{document}